\documentclass[journal, twocolumn, a4paper]{IEEEtran}

\usepackage{cite}
\usepackage{url}
\usepackage{graphicx}
\usepackage{amssymb,amsmath,bm}
\usepackage{textcomp}
\usepackage{booktabs}
\usepackage{hyperref}
\usepackage{multirow}
\usepackage{tabularx}
\usepackage{verbatim}
\usepackage[table,xcdraw]{xcolor}
\newcolumntype{Y}{>{\centering\arraybackslash}X}

\begin{document}

\title{Quasi-Periodic WaveNet: An Autoregressive Raw Waveform Generative Model with Pitch-dependent Dilated Convolution Neural Network}

\author{Yi-Chiao Wu,
        Tomoki Hayashi,
        Patrick Lumban Tobing,~\IEEEmembership{Member,~IEEE,}
        Kazuhiro Kobayashi,~\IEEEmembership{Member,~IEEE,}\\
        and~Tomoki Toda,~\IEEEmembership{Member,~IEEE}
\thanks{Manuscript received xxx xx, 2019; revised xxx xx, 2020. This work was supported in part by theJapan Science and Technology Agency (JST), Precursory Research for Embryonic Science and Technology (PRESTO) under Grant JPMJPR1657, in part by the JST, CREST under Grant JPMJCR19A3, and in part by the Japan Society for the Promotion of Science (JSPS) Grants-in-Aid for Scientific Research (KAKENHI) under Grant 17H06101.}
\thanks{Y.-C. Wu and P. L. Tobing are with Graduate School of Informatics, Nagoya University, Aichi, Japan (e-mail: \{yichiao.wu, patrick.lumbantobing\}@g.sp.m.is.nagoya-u.ac.jp)}%
\thanks{T. Hayashi is with Graduate School of Information Science, Nagoya University, Aichi, Japan (e-mail: hayashi.tomoki@g.sp.m.is.nagoya-u.ac.jp)}
\thanks{K. Kobayashi is with Information Technology Center, Nagoya University, Aichi, Japan (e-mail: kobayashi.kazuhiro@g.sp.m.is.nagoya-u.ac.jp)}
\thanks{T. Toda is with Information Technology Center, Nagoya University, Aichi, Japan (e-mail: tomoki@icts.nagoya-u.ac.jp)}
}

\markboth{Journal of \LaTeX\ Class Files,~Vol.~0, No.~0, July~2020}%
{Shell \MakeLowercase{\textit{et al.}}: Bare Demo of IEEEtran.cls for IEEE Journals}

\maketitle

\begin{abstract}
In this paper, a pitch-adaptive waveform generative model named Quasi-Periodic WaveNet (QPNet) is proposed to improve the limited pitch controllability of vanilla WaveNet (WN) using pitch-dependent dilated convolution neural networks (PDCNNs). Specifically, as a probabilistic autoregressive generation model with stacked dilated convolution layers, WN achieves high-fidelity audio waveform generation. However, the pure-data-driven nature and the lack of prior knowledge of audio signals degrade the pitch controllability of WN. For instance, it is difficult for WN to precisely generate the periodic components of audio signals when the given auxiliary fundamental frequency ($F_{0}$) features are outside the $F_{0}$ range observed in the training data. To address this problem, QPNet with two novel designs is proposed. First, the PDCNN component is applied to dynamically change the network architecture of WN according to the given auxiliary $F_{0}$ features. Second, a cascaded network structure is utilized to simultaneously model the long- and short-term dependencies of quasi-periodic signals such as speech. The performances of single-tone sinusoid and speech generations are evaluated. The experimental results show the effectiveness of the PDCNNs for unseen auxiliary $F_{0}$ features and the effectiveness of the cascaded structure for speech generation.
\end{abstract}


\section{Introduction}

\IEEEPARstart{R}{aw} waveform generation of audio signals like speech and music is a commonly used technique as the core of many applications such as text-to-speech (TTS), voice conversion (VC), and music synthesis. However, because of the extremely high temporal resolution (sampling rates are usually higher than 16~kHz) and the very long term dependence of audio signals, directly modeling the raw waveform signals is challenging. To overcome these difficulties, in conventional synthesis techniques, audio signals are usually encoded into low temporal resolution acoustic features and then audio waveforms are decoded on the basis of these acoustic features. The analysis-synthesis (encoding-decoding) technique is called the vocoder~\cite{vocoder_1939, vocoder_1966, phase_vocoder}, which is often built on a source-filter~\cite{source_filter} speech production model including source excitations and vocal tracts. However, because of the oversimplified assumptions of the speech generation mechanism imposed on conventional vocoders such as STRAIGHT~\cite{straight} and WORLD~\cite{world}, the lost temporal details and phase information lead to the serious quality degradation of these conventional vocoders.

Owing to the recent development of deep learning, many neural-based audio generation models~\cite{wavenet, samplernn, fftnet, wavernn, lpcnet, pwn, clarinet, waveglow, flowavenet, nsf_2019, nsf_2020, pap_gan} have been proposed to generate raw audio waveforms without the various assumptions imposed on conventional vocoders. That is, advanced, and deep network architectures directly model the long-term dependence of high-temporal-resolution audio waveforms. In this paper, we focus on WaveNet (WN)~\cite{wavenet}, which is one of the state-of-the-art audio generation models and has been applied to a variety of applications such as music generation~\cite{wn_music}, text-to-speech (TTS)~\cite{deepvoice3, tacotron2}, speech coding~\cite{wn_code}, speech enhancement~\cite{wn_se_b, wn_se}, and voice conversion (VC)~\cite{vc_wn_2017, nu_p_2018, nu_np_2018, cl_2018, cl_2020}. The main core of WN is an autoregressive (AR) network modeling the probability distribution of each audio sample conditioned on auxiliary features and a specific number of previous samples called a {\it receptive field}. To handle the very long term dependence of audio signals, a stacked dilated convolution network (DCNN)~\cite{dcnn} structure is utilized to efficiently extend the {\it receptive field}. Furthermore, the WN vocoder~\cite{sd_wn_vocoder, si_wn_vocoder, ns_wn_vocoder, sp_wn_vocoder}, which conditions WN on the acoustic features extracted by conventional vocoders to recover the lost information, achieves significant speech quality improvements for speech generation by replacing the synthesis process of traditional vocoders.

Although WN attains excellent performance in high-fidelity speech generation, the fixed architecture is inefficient and the lack of prior audio-related knowledge limits the pitch controllability of the WN vocoder. Specifically, because of the quasi-periodicity of speech, each sample may have a specific dependent field related to its periodicity instead of a fixed {\it receptive field} that presumably includes many redundant previous samples. The requirement of a long {\it receptive field} for modeling speech dependency will lead to a huge network and high demands for computation power. The data-driven architecture without prior speech knowledge only implicitly models the relationship between the periodicity of waveform signals and the auxiliary fundamental frequency ($F_{0}$) features, which may not explicitly generate speech with the precise pitch corresponding to the auxiliary $F_{0}$ values, especially in an unseen $F_{0}$ case. However, pitch controllability is an essential feature for the definition of a vocoder.

To address these problems, inspired by the source-filter model~\cite{source_filter} and code-excited linear prediction (CELP) codec~\cite{celp_1984, celp_1985}, we propose Quasi-Periodic WaveNet (QPNet)~\cite{qpnet_2019, qpnet_vc} with a pitch-dependent dilated convolution neural network (PDCNN). Specifically, the generation process of periodic signals can be modeled as the generation of a single pitch cycle signal (short-term correlation) and then extending this single cycle signal to form the whole periodic sequences on the basis of pitches (long-term correlation). As a result, we develop QPNet including two cascaded WNs with different DCNNs. Vanilla WN with fixed DCNNs is the first stage, which is used to model the relationship between the current sample and a specific segment of the nearest previous samples, and the second stage utilizes the PDCNNs to link the correlations of the relevant segments in the current and previous cycles. The Pitch-adaptive architecture allows each sample to have an exclusive {\it receptive field} length corresponding to the auxiliary $F_{0}$ features and improves the pitch controllability by introducing the periodicity information into the network. The proposed QPNet with the improved pitch controllability is more in line with the definition of a vocoder. Furthermore, a more compact network size while achieving acceptable quality similar to that of vanilla WN is feasible for QPNet because of the more efficient way the {\it receptive field} is extended, which is highly related to the modeling capability.

The paper is organized as follows. In Section~\ref{related_works}, we review the speech manipulation of STRAIGHT and WORLD and the development of recent neural-based speech generation models. In Section~\ref{wavenet}, a brief introduction to WN is presented. In Section~\ref{qpnet}, we describe the concepts and details of QPNet. In Sections~\ref{test_1} and~\ref{test_2}, we report objective and subjective experimental results to evaluate the effectiveness of QPNet for generating high-temporal-resolution periodic sinusoid signals and quasi-periodic speech, respectively. Finally, the conclusion is given in Section~\ref{conclusion}.

\section{Related Work} \label{related_works}

\subsection{Speech Manipulation of STRAIGHT and WORLD}
The human speech production process is usually described as a source-filter mode~\cite{source_filter}. An excitation (source) signal is first generated by vocal fold movements (for voiced sound) or constriction and closure of specific points along the human vocal tract (for unvoiced and plosive sounds). Then the generated-excitation signal is modulated by the resonance of the vocal and nasal tracts and transferred by the lips. For a discrete-time digital system, the excitation signal is represented as a digital signal, and the spectral properties of the vocal and nasal tracts resonance and the lip radiation are represented as a digital filter. The digital source signal excites the digital filter to generate speech signals. 

To flexibly manipulate speech components such as pitch and timbre, many source-filter vocoder techniques have been proposed. However, the spectral estimation of early approaches such as linear predictive coding (LPC) vocoder~\cite{lpc_1971, lpc_1982} are susceptible to signal periodicity~\cite{lpc_1991}. Specifically, getting a stable spectral envelope regardless of the windowing temporal positions is difficult for the voiced speech analysis. The time-variant pitch and natural fluctuations make the spectral analysis suffer from the periodicity interferences because of the fixed window length. 

To address this problem, STRAIGHT~\cite{straight} and WORLD~\cite{world} have been proposed. The STRAIGHT vocoder adopts a pitch-synchronized mechanism~\cite{pitch_sync} with phasic interference reduction and oversmoothing compensation to extract stable spectra, which are highly uncorrelated to the instantaneous $F_0$. Specifically, when extracting features, the window of each frame has a different length according to the $F_0$ of this frame to avoid the periodicity interferences from the voiced speech. Furthermore, as an improved and real-time version, the WORLD vocoder also adopts the pitch-synchronized concept for its spectral analysis~\cite{cheaptrick}.

Although the STRAIGHT and WORLD vocoders achieve high flexibility of speech manipulation, the lost details and phase information problems cause speech quality degradation. The recent neural vocoders greatly improve speech quality but suffer from the limited flexibility of speech manipulation. As a result, we propose a pitch-adaptive component, PDCNN, and a cascaded structure to improve the pitch controllability of the WN vocoder while trying to keep a similar speech quality. The proposed QPNet is also conditioned on the WORLD-extracted features, and we expect QPNet is capable to manipulate pitch like the WORLD vocoder.

\subsection{Neural Vocoder}
Recent mainstream speech generation techniques use AR models such as WN~\cite{wavenet} and SampleRNN~\cite{samplernn} to model the very long term dependence of speech signals with high temporal resolution. For instance, vanilla WN adopts linguistic and $F_{0}$ features to guide the network to generate desired speech waveforms. However, in contrast to the linguistic and $F_{0}$ auxiliary features, the WN vocoder~\cite{sd_wn_vocoder, si_wn_vocoder, ns_wn_vocoder, sp_wn_vocoder} adopts acoustic features as the auxiliary features for a more efficient training that requires much less training data. Many acoustic features have been applied to these AR vocoders such as the mel-cepstral coefficients ($mcep$) with band aperiodicity ($ap$) and $F_{0}$ features, which are extracted from WORLD~\cite{sd_wn_vocoder, si_wn_vocoder, ns_wn_vocoder} or STRAIGHT~\cite{srnn_vocoder}, and mel-spectrograms with $F_{0}$ features~\cite{sp_wn_vocoder}.

Furthermore, to achieve acceptable speech quality, the basic AR vocoders usually require a huge network for the long {\it receptive field}. However, although the speech qualities of these basic AR vocoders are significantly higher than those of the traditional vocoders, the AR mechanism and the complicated network structure make these AR vocoders difficult to generate speech in real-time~\cite{wavenet, samplernn}. To tackle this issue, the authors of FFTNet~\cite{fftnet} and WaveRNN~\cite{wavernn} proposed more compact AR vocoders with specific network structures based on speech-related knowledge and efficient computation mechanisms. Moreover, AR models generating glottal excitation~\cite{glota_wn_1, glota_wn_2} and linear predictive coding (LPC) residual~\cite{lpcnet} signals have been proposed to ease the burden of modeling speaker identity and spectral information. Because of the speaker-independent characteristic of these source signals, the requirements for the network capacity and speaker adaptation of these glottal vocoders and LPCNet are greatly reduced.

In addition, flow-based~\cite{nice, norm_flow} non-AR vocoders have been proposed for efficient parallel generations. For example, parallel WaveNet~\cite{pwn} and ClariNet~\cite{clarinet} with inverse autoregressive flow (IAF)~\cite{iaf} and WaveGlow~\cite{waveglow} and FloWaveNet~\cite{flowavenet} with Glow~\cite{glow} model an invertible transformation between a simple probability distribution of noise signals and a target distribution of speech signals for generating waveforms from a known noise sequence.

Non-AR vocoders with mixed sine-based excitation inputs produced on the basis of $F_{0}$ and Gaussian noise~\cite{nsf_2019, nsf_2020} or periodic sinusoid signals and aperiodic Gaussian noise inputs~\cite{pap_gan} have also been proposed to simultaneously generate whole waveforms while attaining pitch controllability via the manipulation of the periodic inputs. However, to synchronize the phases of generated and ground truth waveforms during training, these models need a handcrafted design of the input signal or a GAN~\cite{gan} structure, which increases the complexity of the models.

Instead of the carefully designed inputs and specific networks, we proposed a simple module PDCNNs, which can be easily applied to any CNN-based generative model to improve its audio signal modeling capability by introducing pitch information into the network. We applied PDCNNs to WN to develop a pitch-dependent adaptive network QPNet~\cite{qpnet_2019, qpnet_vc} for speech generation with arbitrary $F_{0}$ values. In this paper, we further evaluate the periodical modeling capability of QPNet with PDCNNs for nonspeech sinusoid signals generation and comprehensively explore the effectiveness of the QPNet model with different cascade orders,  network structures, and adaptive dilation sizes.

\section{WaveNet for Speech Generation} \label{wavenet}

\subsection{WaveNet}

Because an audio waveform is a sequential signal with a strong long-term dependency, WN~\cite{wavenet} is used to model audio signals in an AR manner that predicts the distribution of each waveform sample on the basis of its previous samples. The conditional probability function can be formulated as
\begin{align}
P\left ( \boldsymbol{x} \right )=\prod_{t=1}^{T}P\left ( x_t\mid x_{t-1},\ldots,x_{t-r} \right )
\label{eq:uc_prob}
\end{align}
where $t$ is the sample index, $x_t$ is the current audio sample, and $r$ is a specific length of the previous samples called a {\it receptive field}. Instead of the general recurrent structure for AR modeling, WN applies stacked convolution neural networks (CNNs) with a dilated mechanism and a causal structure to model the very long term dependence and causality of audio signals. Since the modeling capability of WN is highly related to the amounts of the previous samples taken into consideration for predicting the current sample, the dilated mechanism improves the efficiency of extending the {\it receptive field}. Moreover, a categorical distribution is applied to model the conditional probability whereas audio signals are encoded into 8 bits by using the $\mu$-law algorithm. The categorical distribution is flexible to model an arbitrary distribution of target speech. Taken together, the data flow of WN is as follows: previous audio samples pass through a causal layer and several residual blocks with DCNNs, gated structures, and residual and skip connections. Specifically, the gated structure for enhancing the modeling capability of the network is formulated as
\begin{align}
\boldsymbol{z}^{(\mathrm{o})}=\tanh\left (V_{f, k} \ast \boldsymbol{z}^{(\mathrm{i})} \right )
\odot \sigma \left (V_{g, k} \ast \boldsymbol{z}^{(\mathrm{i})} \right )
\label{eq:uc_gated}
\end{align}
where $\boldsymbol{z}^{(\mathrm{i})}$ and $\boldsymbol{z}^{(\mathrm{o})}$ are the input and output feature maps of the gated structure, respectively. $\boldsymbol{V}$ is a trainable convolution filter, $\ast$ is the convolution operator, $\odot$ is an element-wise multiplication (Hadamard product) operator, $\sigma$ is a sigmoid function, $k$ is the layer index, and $f$ and $g$ are the filter and gate, respectively. Finally, the summation of all skip connections is processed by two ReLU~\cite{relu} activations with $1\times1$ convolutions and one softmax layer to output the predicted distribution of the current audio sample.

Furthermore, to guide the WN model to generate desired contents, the vanilla WN is conditioned on not only previous samples but also linguistic and $F_0$ features. The conditional probability is modified as
\begin{align}
P\left ( \boldsymbol{x}\mid\boldsymbol{h} \right )
=\prod_{t=1}^{T}P\left ( x_t\mid x_{t-1},\ldots,x_{t-r},\boldsymbol{h} \right )
\label{eq:c_prob}
\end{align}
where $\boldsymbol{h}$ is the vector of the auxiliary features (linguistic and $F_0$ features), and the gated activation with auxiliary features becomes
\begin{align}
\boldsymbol{z}^{(\mathrm{o})}=
&\tanh\left (V_{f, k}^{(\mathrm{1})} \ast \boldsymbol{z}^{(\mathrm{i})} 
+ V_{f, k}^{(\mathrm{2})}\ast \boldsymbol{h}^{\prime} \right )\nonumber \\
&\odot \sigma \left (V_{g, k}^{(\mathrm{1})} \ast \boldsymbol{z}^{(\mathrm{i})}
+ V_{g, k}^{(\mathrm{2})}\ast \boldsymbol{h}^{\prime} \right )
\label{eq:c_gated}
\end{align}
where $V^{(\mathrm{1})}$ and $V^{(\mathrm{2})}$ are trainable convolution filters, and ${h}^{\prime}$ is the temporal extended auxiliary features, whose temporal resolution matches to the speech samples.

\subsection{WaveNet Vocoder}
Many conventional vocoders~\cite{straight, world} are built on the basis of a source-filter architecture~\cite{source_filter}, which models the speech generation process as a spectral filter driven by the source excitation signal. However, the oversimplified assumptions, such as time-invariant linear filters and stationary Gaussian processing make the vocoders lose some essential information of speech such as phase and temporal details, and it causes marked quality degradation. To address this problem, the authors of~\cite{sd_wn_vocoder, si_wn_vocoder} proposed the WN vocoder, which conditions WN on the auxiliary acoustic features extracted by a conventional vocoder to generate raw speech waveforms. That is, the WN vocoder replaces the synthesis part of conventional vocoders to synthesize high-fidelity speech on the basis of the prosodic and spectral acoustic features extracted by conventional vocoders. Furthermore, conditioning WN on the acoustic features greatly reduces the requirements of the amounts of the training data, and it makes WN more tractable.

\subsection{Problems in Using WaveNet as A Vocoder}
As a vocoder, WN achieves high speech quality, but it lacks pitch controllability, which is an essential feature of conventional vocoders. Specifically, the WN vocoder has difficulties in generating speech with precise pitch conditioning on the $F_0$ values that are not observed in the $F_0$ range of training data~\cite{qpnet_2019}. Even though the $F_0$ and spectral features are within the observed range, an unseen combination of the auxiliary features still markedly degrades the generation performance of the WN vocoder~\cite{vc_wn_2017, nu_np_2018, nu_p_2018, cl_2018, cl_2020}. The possible reasons for this problem are that WN lacks prior speech knowledge and does not explicitly model the relationship between the auxiliary $F_0$ feature and pitch. The defect makes the WN vocoder inconsistent with the definition of a vocoder. Moreover, since the fixed WN architecture assumes each sample has the same length of the {\it receptive field}, the inefficient {\it receptive field} extending may lead to the costly requirements of a huge network and lots of computation power.

\begin{figure}[t]
\centering
\centerline{\includegraphics[width=1.0\columnwidth]{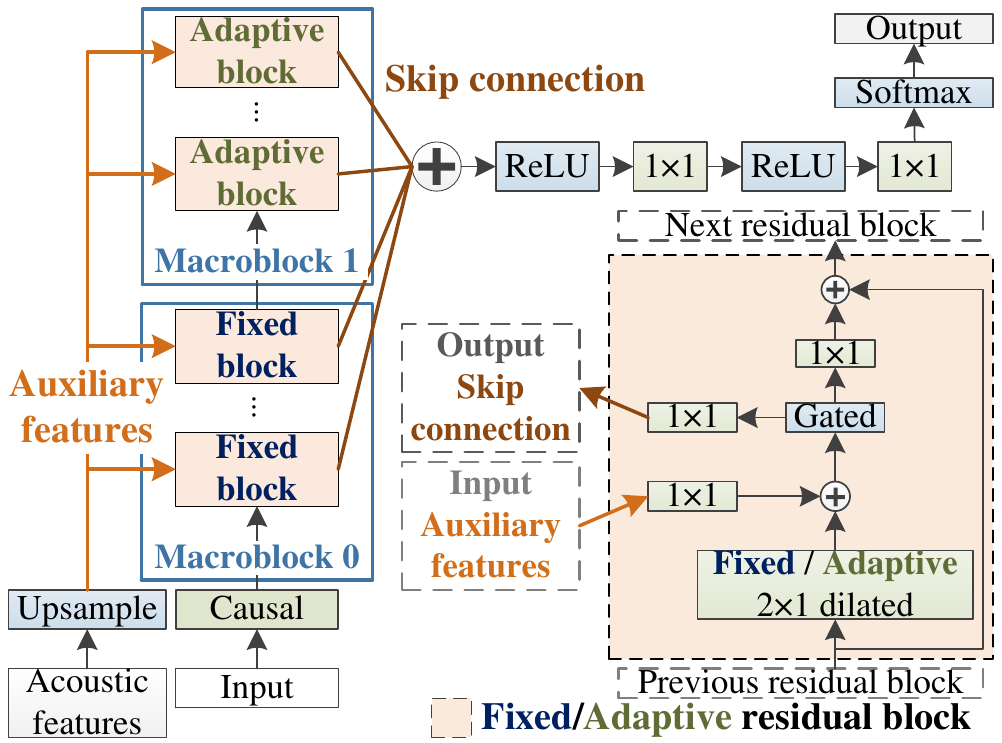}}
\caption{Quasi-Periodic WaveNet vocoder architecture.}
\label{fig:qpnet}
\end{figure}

\section{Quasi-Periodic WaveNet} \label{qpnet}

To improve the efficiency of {\it receptive field} extension and pitch controllability, QPNet introduces the prior pitch information into WN by dynamically changing the network structure according to the auxiliary $F_0$ features. As shown in Fig.~\ref{fig:qpnet}, the main differences between WN and QPNet are the pitch-dependent dilated convolution mechanism handling the periodicity of audio signals and the cascaded structures simultaneously modeling the long- and short-term correlations. The pitch filtering in CELP, which is the basis of the PDCNN, and the details of QPNet are described as follows. 

\begin{figure}[t]
\centering
\centerline{\includegraphics[width=1.0\columnwidth]{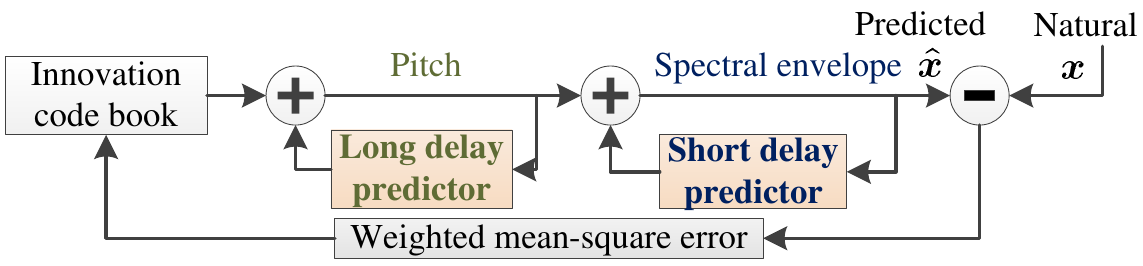}}
\caption{Code-excited linear prediction system.}
\label{fig:celp}
{\ }
\centering
\centerline{\includegraphics[width=1.0\columnwidth]{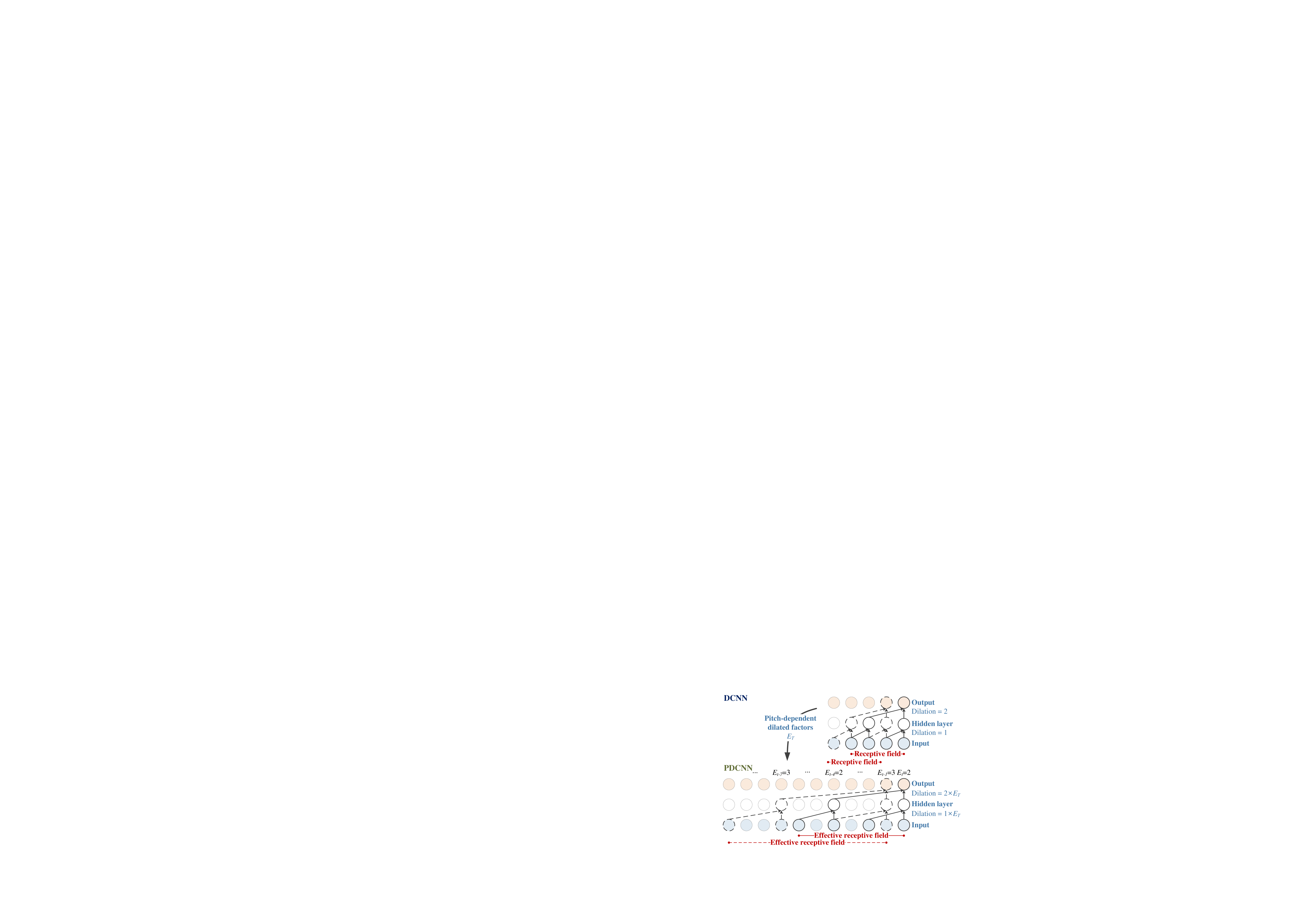}}
\caption{Fixed and pitch-dependent dilated convolution.}
\label{fig:pdcnn}
\end{figure}

\subsection{Pitch Filtering in CELP}
Fig.~\ref{fig:celp} shows a flowchart of the CELP system~\cite{celp_1985}, which includes an innovation signal codebook and two cascaded time-varying linear recursive filters. First, each innovation signal in the codebook is scaled and passed to the pitch filter (long delay) to generate the pitch periodicity of the speech, and then the linear-prediction filter (short delay) restores the spectral envelope to obtain the synthesized speech. Secondly, the mean-square errors between the original and synthesized speech signals are weighted by a linear filter to attenuate/amplify frequency components that are less/more perceptually important. Finally, the optimum innovation signal and the scaled factor are determined by minimizing the weighted mean-square error. To be more specific, the pitch-filtering process can be formulated as
\begin{align}
{c}_{t}^{(\mathrm{o})}=g \times {c}_{t}^{(\mathrm{i})} + b \times {c}_{t-t_d}^{(\mathrm{o})}
\label{eq:celp}
\end{align}
where ${c}^{(\mathrm{i})}$ is the input, ${c}^{(\mathrm{o})}$ is the output, $t_d$ is the pitch delay, $g$ is the gain, and $b$ is the pitch filter coefficient. This periodic feedback structure handling the periodicity of signals is the basis of the proposed PDCNN, and the cascaded recursive structure modeling the hierarchical correlations is also applied to QPNet.

\subsection{Pitch-dependent Dilated Convolution}
The main idea of the PDCNN is that since audio signals have the quasi-periodic property, the network architecture can be dynamically adapted using the prior pitch information. Specifically, the dilated convolution can be formulated as
\begin{align}
 \boldsymbol{y}_{t}^{(\mathrm{o})}
=\boldsymbol{W}^{(\mathrm{c})}\times\boldsymbol{y}_{t}^{(\mathrm{i})}
+\boldsymbol{W}^{(\mathrm{p})}\times\boldsymbol{y}_{t-d}^{(\mathrm{i})},
\label{eq:dcnn}
\end{align}
where $\boldsymbol{y}_{t}^{(\mathrm{o})}$ is the output of the DCNN layer at sample $t$, and $\boldsymbol{y}_{t}^{(\mathrm{i})}$ is the input of the DCNN layer at sample $t$. The trainable $1\times1$ convolution filters $\boldsymbol{W}^{(\mathrm{c})}$ and $\boldsymbol{W}^{(\mathrm{p})}$ are respectively for the current and previous samples. The dilation size $d$ is constant for the vanilla DCNN but time-variant for the PDCNN.

\begin{figure}[t]
\centering
\centerline{\includegraphics[width=1.0\columnwidth]{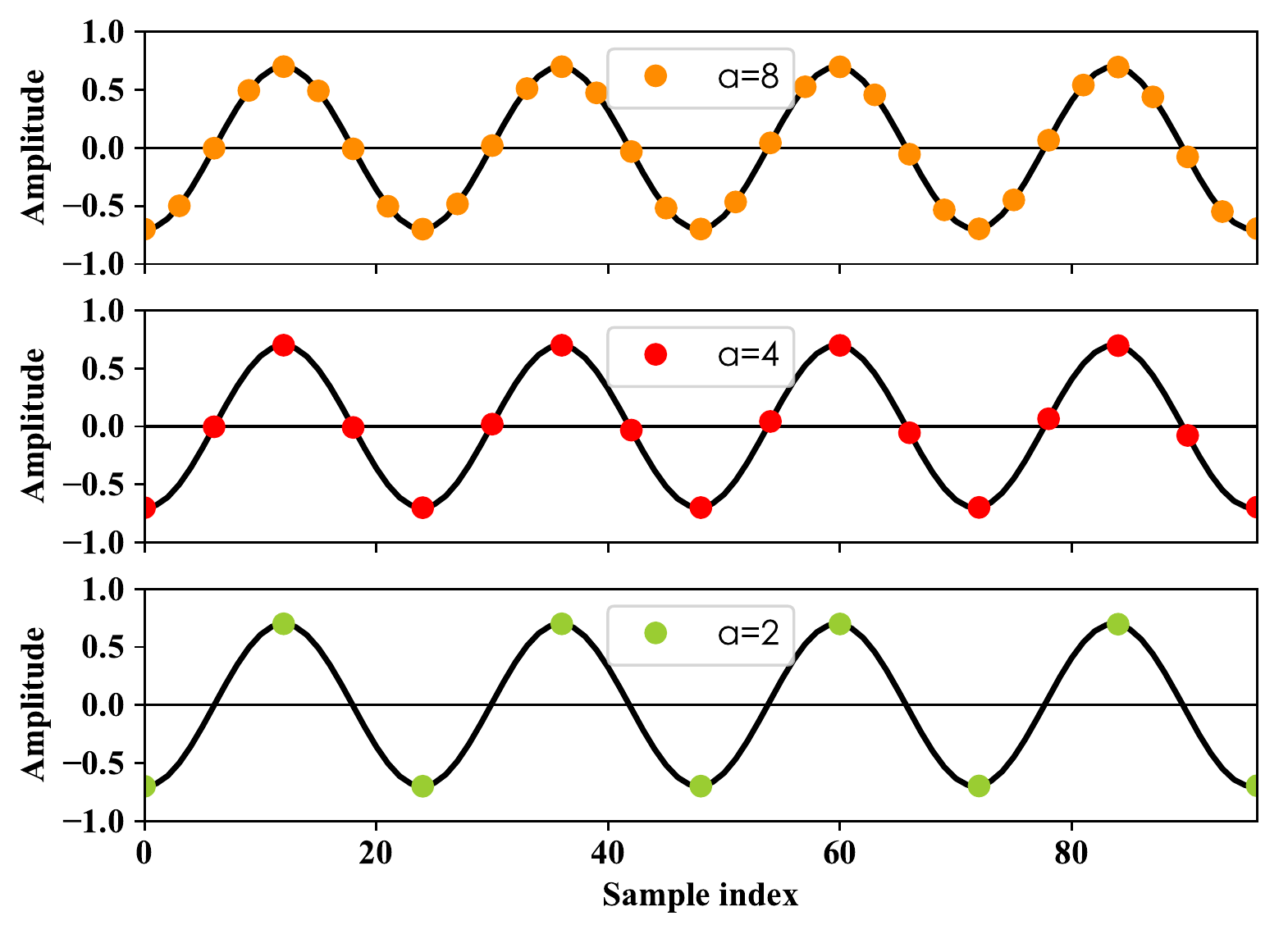}}
\caption{Sampling sparsity of different {\it dense factor} $a$.}
\label{fig:dense}
\end{figure}

To enlarge the {\it receptive field} length, the vanilla WN utilizes stacked chunks including DCNN layers with different dilation sizes. Each chunk contains a specific number of DCNN layers, and each layer (except the first layer) twice the dilation size of the last one. The dilation sizes of the first layers of the chunks are set to one, so the dilation size in each chunk exponentially increases with base two. As shown in Fig.~\ref{fig:pdcnn}, the dilation sizes of PDCNN layers in the stacked adaptive chunks of QPNet follow the same extension rule but multiplied by an extra dilated factor to match the instantaneous pitch of the current sample. The pitch-dependent dilated factor $E_t$ is derived from
\begin{align}
E_{t}=F_{s}/(F_{0,t}\times a),
\label{eq:et}
\end{align}
where $F_{s}$ is the utterance-wise constant sampling rate, $F_{0,t}$ is the fundamental frequency with speech sample index $t$, and $a$ is a hyperparameter called the {\it dense factor}, which indicates the number of samples in one cycle taken into consideration as shown in Fig.~\ref{fig:dense} when predicting the current sample. 

\begin{figure}[t]
\centering
\centerline{\includegraphics[width=1.0\columnwidth]{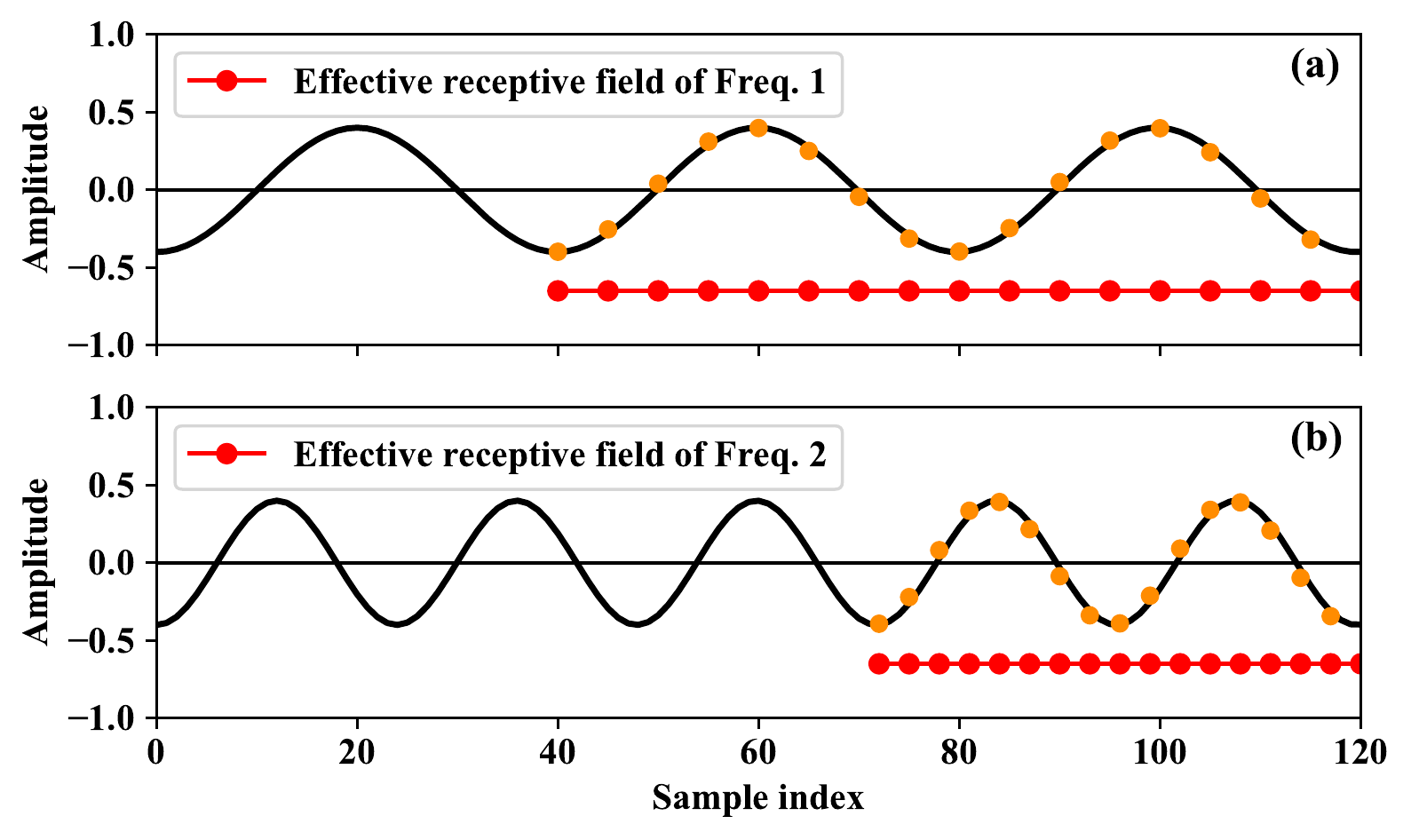}}
\caption{{\it Effective receptive fields} with different $F_0$ values.}
\label{fig:erp_field}
\end{figure}

Specifically, the grid sampling locations of each DCNN is controlled by the dilation size $d$, and the dilation size $d^{\prime}$ of each PDCNN is controlled by the dilated factor $E_t$ as
\begin{align}
d^{\prime}=E_{t} \times d.
\label{eq:dp}
\end{align}
By setting the $F_0$ values and the {\it dense factor} $a$, the network can control the sparsity of the CNN sampling grids to attain the desired {\it effective receptive field} length. As shown in Fig.~\ref{fig:erp_field}, since the sinusoids in Figs.~\ref{fig:erp_field} (a) and (b) have the same {\it dense factors} and sampling rates, even though the frequencies of them are different, the numbers of cycles in their {\it effective receptive fields} are still the same. The difference is the temporal sparsity of the {\it effective receptive field}. That is, fixing the number of sampling grids in each cycle by the {\it dense factor} and changing the gaps between the grid sampling locations by the instantaneous $F_0$ values lead to pitch-dependent and time-variant {\it effective receptive field} lengths.

In summary, the dilated factor $E_t$ is the enlarged ratio of the {\it effective receptive field} length to the {\it receptive field} length, and the ratio of the {\it receptive field} length to the {\it dense factor} $a$ is the number of past cycles in the {\it effective receptive field}. With the pitch-dependent structure, each sample has an exclusive {\it effective receptive field} length, which is efficiently enlarged according to the auxiliary $F_0$ values. In addition, since speech has voiced and unvoiced segments, we have tried to set $E_t$ to one or the value calculated by interpolating the $F_0$ values of the adjacent voiced segments for the unvoiced segments, and the results in Section~\ref{test_2} show that QPNet with the continuous $E_t$ from interpolated $F_0$ values achieves higher speech quality.

\subsection{Cascaded Autoregressive Network}
Most audio signals are sequential and quasi-periodic, so the audio generative models usually simultaneously model the long-term (periodicity) and short-term (aperiodicity) correlations of audio samples. As shown in Fig.~\ref{fig:qpnet}, the proposed QPNet utilizes a cascaded architecture that contains a fixed and an adaptive (pitch-dependent) macroblocks. The fixed macroblock models the sequential relationship between the current sample and a segment of the most recent samples. The adaptive macroblock models the periodic correlations of the current and related past segments in the successive cycles. Specifically, the fixed macroblock (macroblock 0 in Fig.~\ref{fig:qpnet}) of the QPNet is composed of several fixed chunks. Each fixed chunk consists of several stacked residual blocks with DCNNs (fixed blocks), conditional auxiliary features, gated activations, and residual and skip connections, similarly to the vanilla WN. The adaptive macroblock (macroblock 1 in Fig.~\ref{fig:qpnet}) also contains several adaptive chunks, which also have similar stacked residual blocks but with PDCNNs (adaptive blocks). In summary, the cascaded structure of QPNet presumably mimics a similar generative procedure of CELP for quasi-periodic audio signals generation.

\section{Periodic Signal Generation Evaluation} \label{test_1}

\begin{table}[t]
\fontsize{9pt}{10.8pt}
\selectfont
\caption{Architecture of Sinusoidal Generative Model}
\label{tb:model1}
{%
\begin{tabularx}{\columnwidth}{@{}p{2.5cm}YYYY@{}}
\toprule
               & WNf & WNc & (r)QPNet & pQPNet \\ \midrule
Fixed chunk    & 3   & 4   & 3        & -      \\                
Fixed block    & 10  & 4   & 4        & -      \\
Adaptive chunk & -   & -   & 1        & 4      \\
Adaptive block & -   & -   & 4        & 4      \\
CNN$^1$ channel   & \multicolumn{4}{c}{128}       \\
CNN$^2$ channel   & \multicolumn{4}{c}{128}       \\
CNN$^3$ channel   & \multicolumn{4}{c}{64}        \\
Size ($\times 10^6$)     & 2.4 & 1.5 & 1.5      & 1.5    \\ \bottomrule
\end{tabularx}%
}
\\$^1$Causal and dilated CNN 
\\$^2 1 \times 1$ CNN in residual block
\\$^3 1 \times 1$ CNN in output layer
\end{table}

To evaluate pitch controllability of the proposed QPNet with the PDCNNs, we first evaluated the generation quality of simple periodic but high-temporal-resolution signals. That is, the training data of QPNet were sine waves within a specific frequency range and the corresponding $F_0$ values. In the test phase, QPNet was conditioned on an $F_0$ value and a small piece of the related sine wave for the initial {\it receptive field} to generate sinusoid waveforms. 

\subsection{Model Architecture}
In this section, to evaluate the effectiveness of the PDCNN, we compared three types of QPNet with two types of WN in terms of sine wave generation. Specifically, in addition to the basic QPNet, because a sinusoid is a simple periodic signal that can be modeled well by a pitch-dependent structure, the QPNet model with only adaptive residual blocks (pQPNet) was taken into account. The QPNet model with the reverse order of the fixed and adaptive macroblocks (rQPNet) was also considered. Moreover, a compact-size WN (WNc) and a full-size WN (WNf) models were evaluated as the references.

The details of the network architectures are shown in Table~\ref{tb:model1}. Since the numbers of CNN channels were the same for all models, the model sizes were proportional to the numbers of the chunks and residual blocks. For instance, the WNf contained 3 chunks and each chunk included 10 residual blocks, so the model size of the WNf was larger than that of the WNc, which only had 4 chunks with 4 residual blocks in each chunk. The learning rate was $1 \times 10^{-4}$ without decay, the minibatch size was one, the batch length was 22,050 samples, the training epochs were two, and the optimizer was Adam~\cite{adam} for all models.

\subsection{Evaluation Setting}
Because the pitch range of most speech is around 80--400~Hz, the training sine waves were set to be in the same range with a step size of 20~Hz (ex: 80, 100, 120 … Hz). Each model had a related one-dimensional $F_0$ value as its auxiliary feature. Since the single-tone generation was evaluated, the auxiliary features of all samples in one utterance were the same. To prevent the networks from suboptimal training and lacking the generality for sinusoid generations with unseen $F_0$ values, both sinusoid and auxiliary signals were mixed with white noise.

The signal-to-noise ratio (SNR) of the sine waves was around 20~dB, and the noise of the auxiliary feature was a random sequence between -1 and 1. Random initial phases were also applied to the sinusoid signals. The number of training utterances was 4000, and each utterance was one second. The ground truths were clean sinusoid signals, so each model was trained as a denoising network. The test data included 20 different $F_0$ values, which were 10--80~Hz with a step size of 10~Hz, 100--400~Hz with a step size of 100~Hz, and 450--800~Hz with a step size of 50~Hz, and each $F_0$ value contained 10 test utterances with different phase shifts. Both training and test data were encoded using the $\mu$-law into 8~bits, and the sampling rate was 22,050~Hz.

In the test stage, the initial {\it receptive field} of each network was fed with the noisy test sine wave, and the length of the generated sinusoid was set to 1s. The test data were divided into 10--40~Hz (under $1/2 L$), 50--80~Hz (above $1/2 L$), 100--400~Hz (inside), 450-–600~Hz (under $3/2 U$), and 650–-800~Hz (above $3/2 U$) subsets. $L$ is the lower bound and $U$ is the upper bound of the inside $F_0$ range, which was the $F_0$ range of the training data. As a result, the under $1/2 L$ and above $1/2 L$ $F_0$ ranges are the lower outside $F_0$ range, and the under $3/2 U$ and above $3/2 U$ $F_0$ ranges are the higher outside $F_0$ range.

\subsection{Performance Measurement}
The quality of each generated waveform was evaluated on the basis of the SNR and the root-mean-square error (RMSE) of the log $F_0$ value measured from the peak of the power spectral density (PSD). Specifically, because the SNRs are related to the noisy degrees of the generated signals, the SNR values will indicate the generated signals are clear sinusoids or not. Since it was a single-tone sinusoid generation test, the high log $F_0$ RMSEs might imply that the generated signals include much harmonic noise or the frequencies of these signals are incorrect. In other words, the generated signal with a high SNR and a high RMSE might be a clear sinusoid with an inaccurate frequency,  the generated signal with a low SNR and a high RMSE might be a noisy sinusoid with much harmonic noise, and the generated signal with a very low SNR might be a noise-like signal.

\subsection{Dense Factor}
To explore the efficient {\it dense factor} value of the PDCNNs, the sinusoid generative qualities of the pQPNet models with different {\it dense factors} were evaluated. Since the chunk and block numbers of the pQPNets were set to four, the length of the {\it receptive fields} was 61 samples. That is, the {\it receptive fields} included from 61 past cycles to less than one cycle according to the {\it dense factors} from $2^0$ to $2^6$. Moreover, in contrast to containing a fixed number of past cycles for sinusoids with arbitrary pitch, the {\it receptive fields} of the WNf contained 11 past cycles for 80~Hz sinusoids and 56 past cycles for 400~Hz sinusoids when the sampling rate was 22,050~Hz. As a result, the {\it effective receptive fields} of the pQPNet with a {\it dense factor} 2 already contained a comparative number of the past cycles as the WNf. Since the pQPNets introduced prior pitch knowledge into the network, the required number of the past cycles for modeling the sinusoids might be less than that of the WNf.

\begin{table}[t]
\fontsize{9pt}{10.8pt}
\selectfont
\caption{SNR ({\rm dB}) of Sinusoid Generation with Different Dense Factors}
\label{tb:dense_snr}
{%
\begin{tabular}{@{}lccccccc@{}}
\toprule
Dense $a$      & $2^0$ & $2^1$ & $2^2$ & $2^3$ & $2^4$ & $2^5$ & $2^6$   \\ \midrule
Under $1/2 L$  & 6.7  & 14.4 & 20.8 & 21.9 & 25.8 & \textbf{28.0} & 27.9  \\
Above $1/2 L$  & 19.8 & 11.9 & 21.5 & 26.6 & 24.5 & \textbf{28.9} & 26.4  \\
Inside         & 17.1 & 19.1 & 19.4 & 26.0 & \textbf{29.9} & 23.2 & 17.5  \\
Under $3/2 U$  & 1.1  & 6.7  & 3.0  & 19.9 & \textbf{23.2} & 17.1 & -17.7 \\
Above $3/2 U$  & -8.1 & -0.8 & -0.3 & 2.7  & \textbf{8.3}  & 3.0  & -23.5 \\ \midrule
Average        & 7.3  & 10.3 & 12.9 & 19.4 & \textbf{22.3} & 20.0 & 6.1   \\ \bottomrule
\end{tabular}%
}
\end{table}

\begin{table}[t]
\fontsize{9pt}{10.8pt}
\selectfont
\caption{Log $F_0$ RMSE of Sinusoid Generation \protect \\ with Different Dense Factors}
\label{tb:dense_f0}
{%
\begin{tabular}{@{}lccccccc@{}}
\toprule
Dense $a$      & $2^0$ & $2^1$ & $2^2$ & $2^3$ & $2^4$ & $2^5$ & $2^6$   \\ \midrule
Under $1/2 L$  & 0.26 & \textbf{0.00} & \textbf{0.00} & \textbf{0.00} & 0.03 & 0.05 & 0.14 \\
Above $1/2 L$  & 0.00 & 0.01 & \textbf{0.00} & \textbf{0.00} & 0.01 & 0.01 & 0.10 \\
Inside         & 0.42 & \textbf{0.00} & \textbf{0.00} & 0.01 & 0.01 & 0.02 & 0.03 \\
Under $3/2 U$  & 1.95 & 0.08 & \textbf{0.03} & 0.04 & 0.08 & 0.09 & 0.89 \\
Above $3/2 U$  & 0.61 & \textbf{0.04} & 0.05 & 0.06 & 0.09 & 0.15 & 1.97 \\ \midrule
Average        & 0.65 & 0.03 & \textbf{0.02} & \textbf{0.02} & 0.04 & 0.06 & 0.63   \\ \bottomrule
\end{tabular}%
}
\end{table}

The number of training epochs of the pQPNet models with {\it dense factors} from $2^2$ to $2^6$ was two. For {\it dense factors} of $2^0$ and $2^1$, pQPNet required at least 10 training epochs to attain stable results. As shown in Tables~\ref{tb:dense_snr} and~\ref{tb:dense_f0}, the network with the {\it dense factor} of $2^0$ was very unstable even when already trained with 10 epochs. The results indicate that although the small {\it dense factor} made the network have a long {\it effective receptive field}, the overbrief information of each past cycle might make it difficult to model signals well. For the inside and lower outside $F_0$ ranges, the networks with {\it dense factors} greater than $2^1$ achieved high SNR values. However, the performance of the network with a {\it dense factor} of $2^6$ markedly degraded when the auxiliary $F_0$ values were in the higher outside $F_0$ range. The possible reason is that the PDCNNs of the network degenerated to DCNNs because the $E_t$ became one when the {\it dense factor} was $2^6$ and the $F_0$ values were higher than 350~Hz. Moreover, the log $F_0$ RMSE results show a similar tendency to the SNR results. The networks with {\it dense factors} of $2^0$ and $2^6$ achieved the lowest pitch accuracies while the networks with {\it dense factors} of $2^2$ and $2^3$ achieved the highest pitch accuracies.

Furthermore, according to the Nyquist–Shannon sampling theorem~\cite{nyquist}, a signal can be perfect reconstructed if the bandwidth of the signal is less than the halved sampling rate. Therefore, the {\it dense factor} $2^1$ is theoretically enough to model the periodic signals. The instability and markedly high RMSE results of the pQPNet with {\it dense factor} $2^0$ also confirm this theory. However, in signal processing, oversampling usually improves resolution and SNR, and relaxes filter performance requirements to avoid aliasing. The higher SNR and lower RMSE of the pQPNets with {\it dense factor} $2^2$ and $2^3$ have shown this tendency, and the performance degradation of the pQPNet with {\it dense factor} $2^6$ is caused by the PDCNN degeneration issue, which is irrelevant to the sampling theorem.

In conclusion, the PDCNN with an appropriate {\it dense factor} was found to be robust against the conditions in the outside $F_0$ range, especially in the lower outside $F_0$ range conditions. For the higher outside $F_0$ range conditions, the networks still had acceptable quality until the $F_0$ value exceeded 600~Hz. Therefore, we set the {\it dense factors} to $2^3$ for the models in the following evaluations because of the balance between the generative performance and the number of past cycles covered in its {\it receptive fields}.

\begin{table}[t]
\fontsize{9pt}{10.8pt}
\selectfont
\caption{SNR ({\rm dB}) of Sinusoid Generation with Different Models}
\label{tb:model_snr}
{%
\begin{tabularx}{\columnwidth}{@{}p{1.7cm}YYYYY@{}}
\toprule
            & WNc   & WNf  & pQPNet & QPNet & rQPNet \\ \midrule
Under $1/2 L$ & -18.1 & \textbf{24.3} & 21.9   & -8.1  & 18.4   \\
Above $1/2 L$ & 8.1   & 23.0 & 26.6   & 28.2  & \textbf{28.7}   \\
Inside        & 28.8  & \textbf{34.5} & 26.0   & 25.9  & 27.0   \\
Under $3/2 U$ & 13.7  & 17.6 & \textbf{19.9}   & 8.7   & 19.3   \\
Above $3/2 U$ & -14.1 & -0.4 & \textbf{2.7}    & -18.6 & -8.2   \\ \midrule
Average       & 3.7   & \textbf{19.8} & 19.4   & 7.2   & 17.0   \\ \bottomrule
\end{tabularx}%
}
\end{table}

\begin{table}[t]
\fontsize{9pt}{10.8pt}
\selectfont
\caption{Log $F_0$ RMSE of Sinusoid Generation with Different Models}
\label{tb:model_f0}
{%
\begin{tabularx}{\columnwidth}{@{}p{1.7cm}YYYYY@{}}
\toprule
            & WNc  & WNf  & pQPNet & QPNet & rQPNet \\ \midrule
Under $1/2 L$ & 2.93 & 1.75 & \textbf{0.00}   & 2.00  & 0.18   \\
Above $1/2 L$ & 0.55 & 0.58 & \textbf{0.00}   & 0.02  & \textbf{0.00}   \\
Inside        & 0.01 & \textbf{0.00} & 0.01   & 0.01  & \textbf{0.00}   \\
Under $3/2 U$ & \textbf{0.04} & 0.50 & \textbf{0.04}   & 0.11  & 0.11   \\
Above $3/2 U$ & 0.12 & 0.48 & \textbf{0.06}   & 0.48  & \textbf{0.06}   \\ \midrule
Average       & 0.73 & 0.66 & \textbf{0.02}   & 0.53  & 0.07   \\ \bottomrule
\end{tabularx}%
}
\end{table}

\subsection{Network Comparison}
As shown in Tables~\ref{tb:model_snr} and~\ref{tb:model_f0}, the PDCNNs significantly improved pitch controllability. The PDCNNs made the QP-series networks achieve much higher SNR and lower log $F_0$ RMSE values than the same-size WNc network in both higher and lower outside $F_0$ ranges, and it shows the effectiveness of the PDCNNs to enlarge the {\it effective receptive field} length. Although the full-size WNf attained similar SNRs to the pQPNet, the log $F_0$ RMSE of the WNf was much higher in the outside $F_0$ ranges. The results indicate that the WNf tended to generate the signals in the inside $F_0$ range instead of being consistent with the auxiliary $F_0$ feature. Therefore, the generated waveform of the WNf might still be a perfect sinusoid signal but with an incorrect pitch. The results also imply that the PDCNNs improved the periodical modeling capability using prior pitch knowledge.  

In addition, because of the simple periodic signal generation scenario, the pQPNet with the longest {\it effective receptive fields} and the pure PDCNN structure attained the best generative performance among all QP-series networks. The QPNet and the rQPNet showed some quality degradations when the auxiliary $F_0$ values were far away from the inside $F_0$ range, but they still outperformed the WNc in both measurements and the WNf in terms of log $F_0$ RMSE.

\begin{figure}[t]
\resizebox{1.0\columnwidth}{!}{%
\begin{tabular}{@{}cc@{}}
  \includegraphics{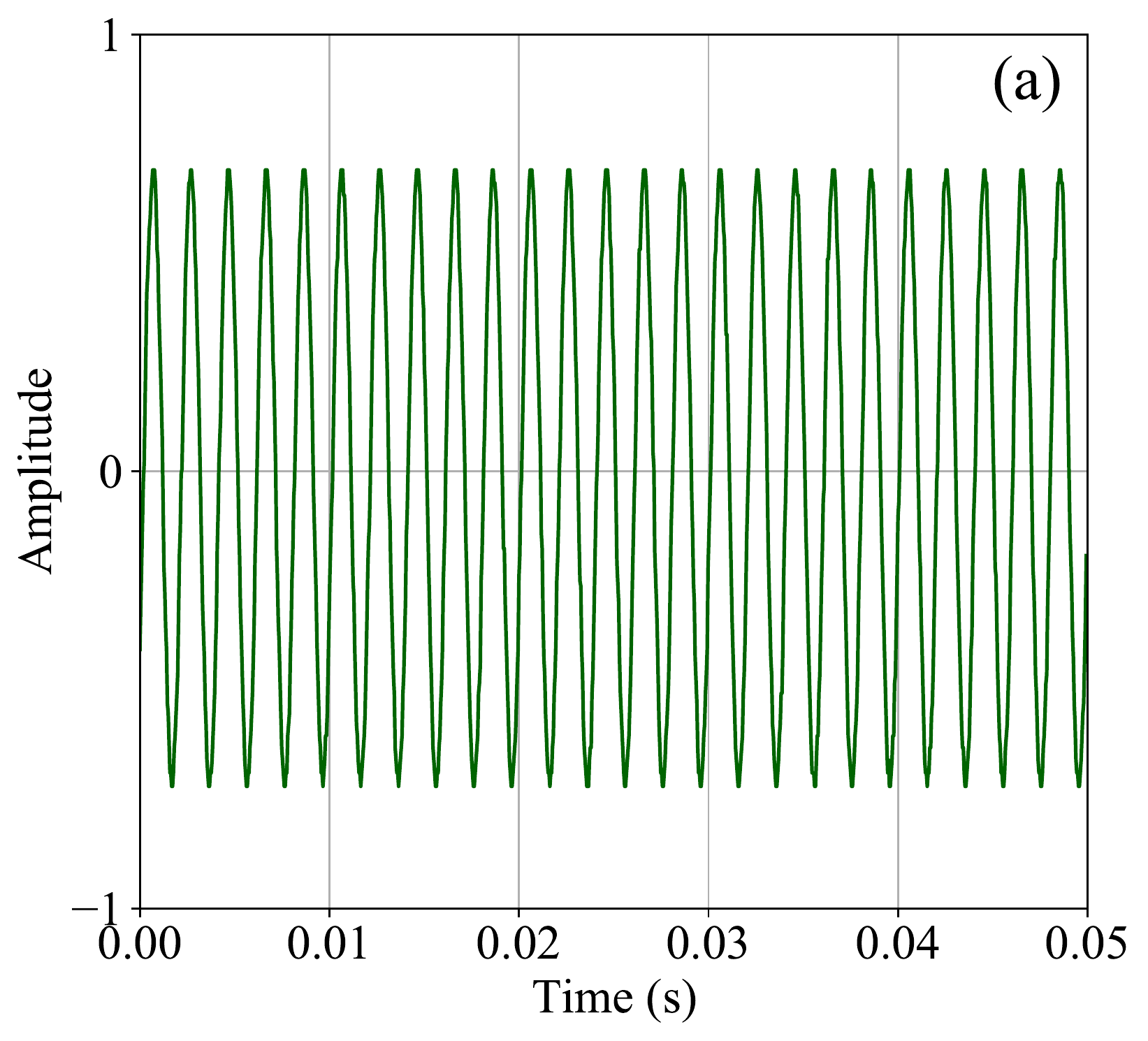}
&  \includegraphics{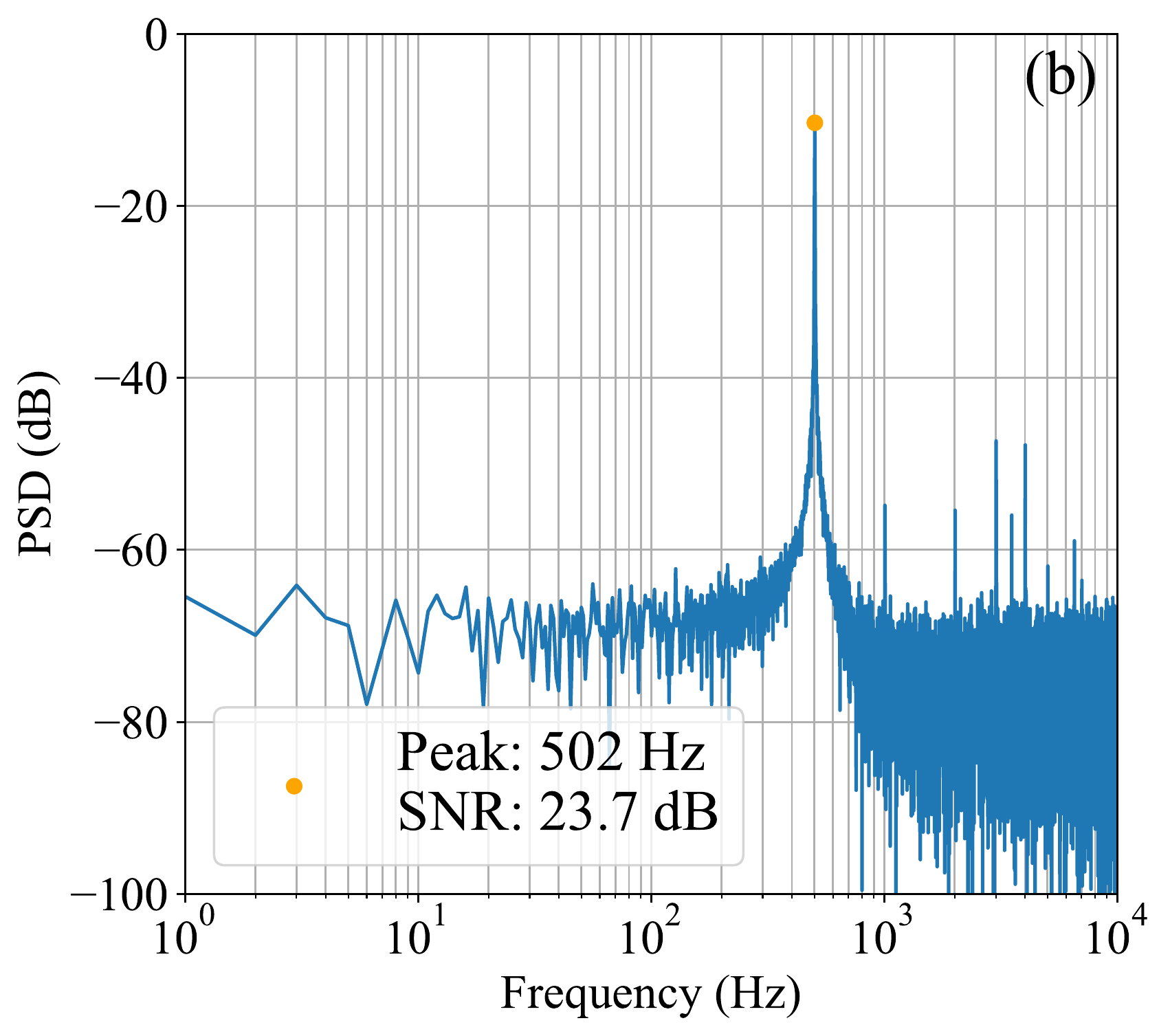} \\
  
  \includegraphics{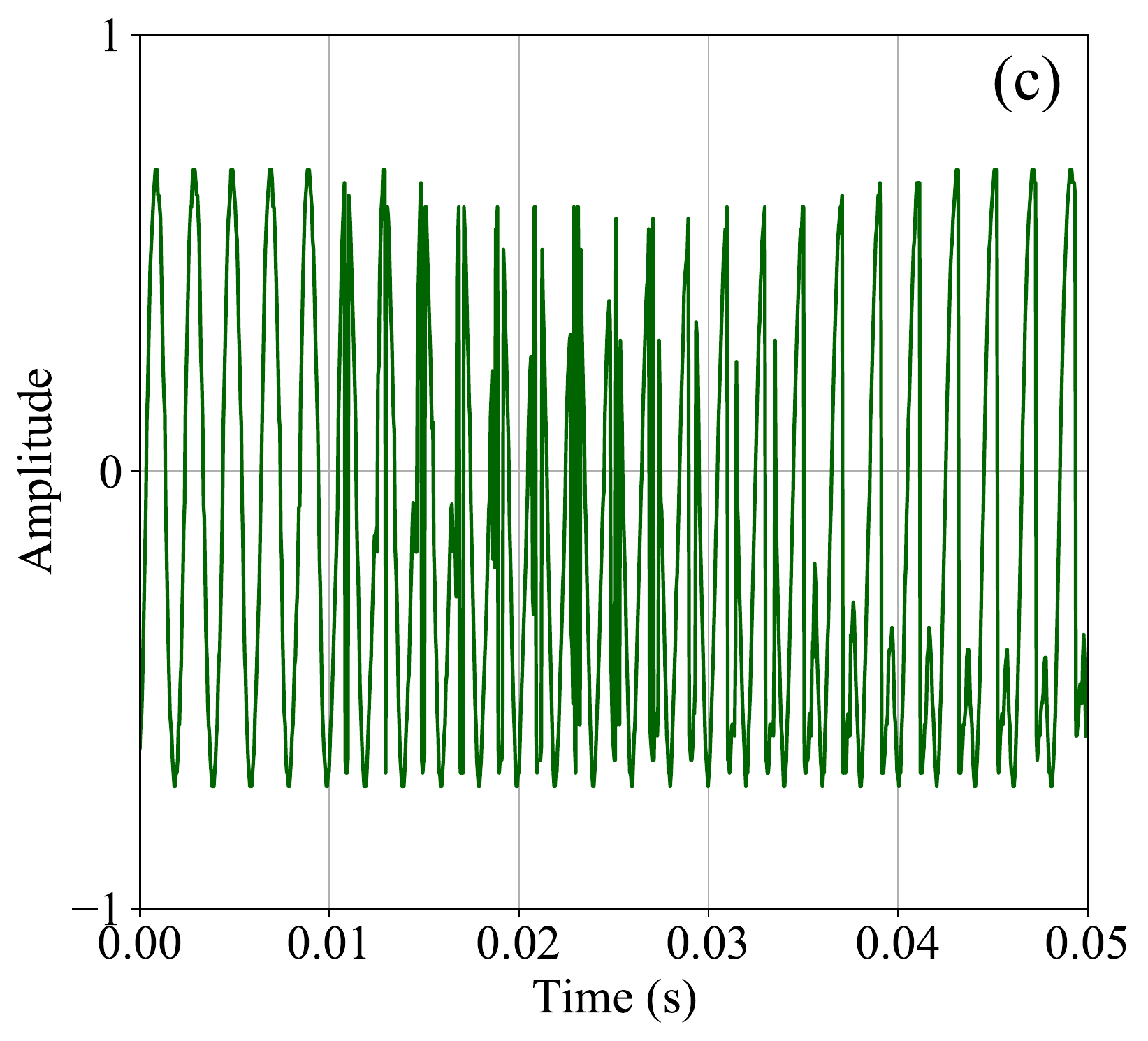}
& \includegraphics{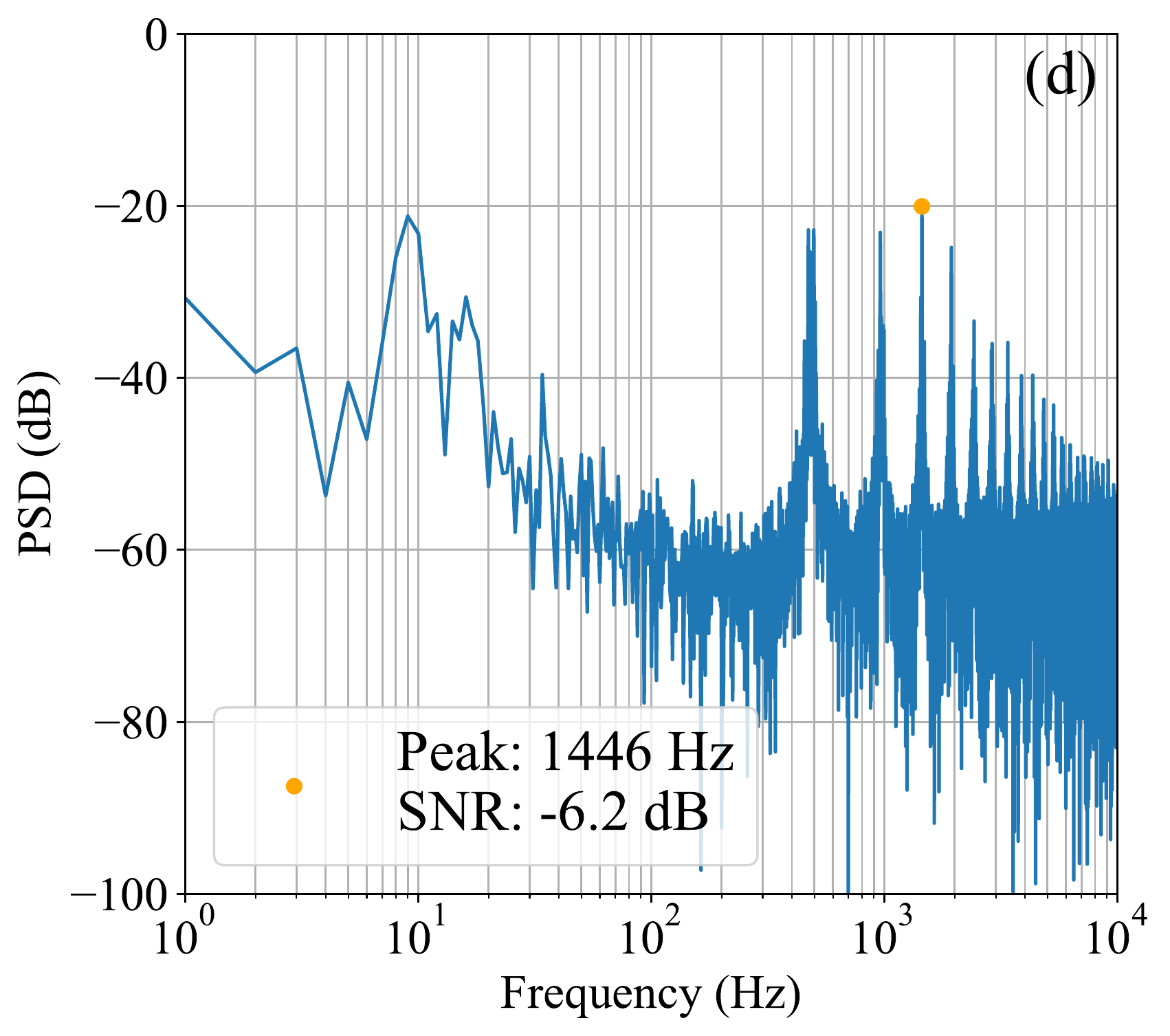} \\

  \includegraphics{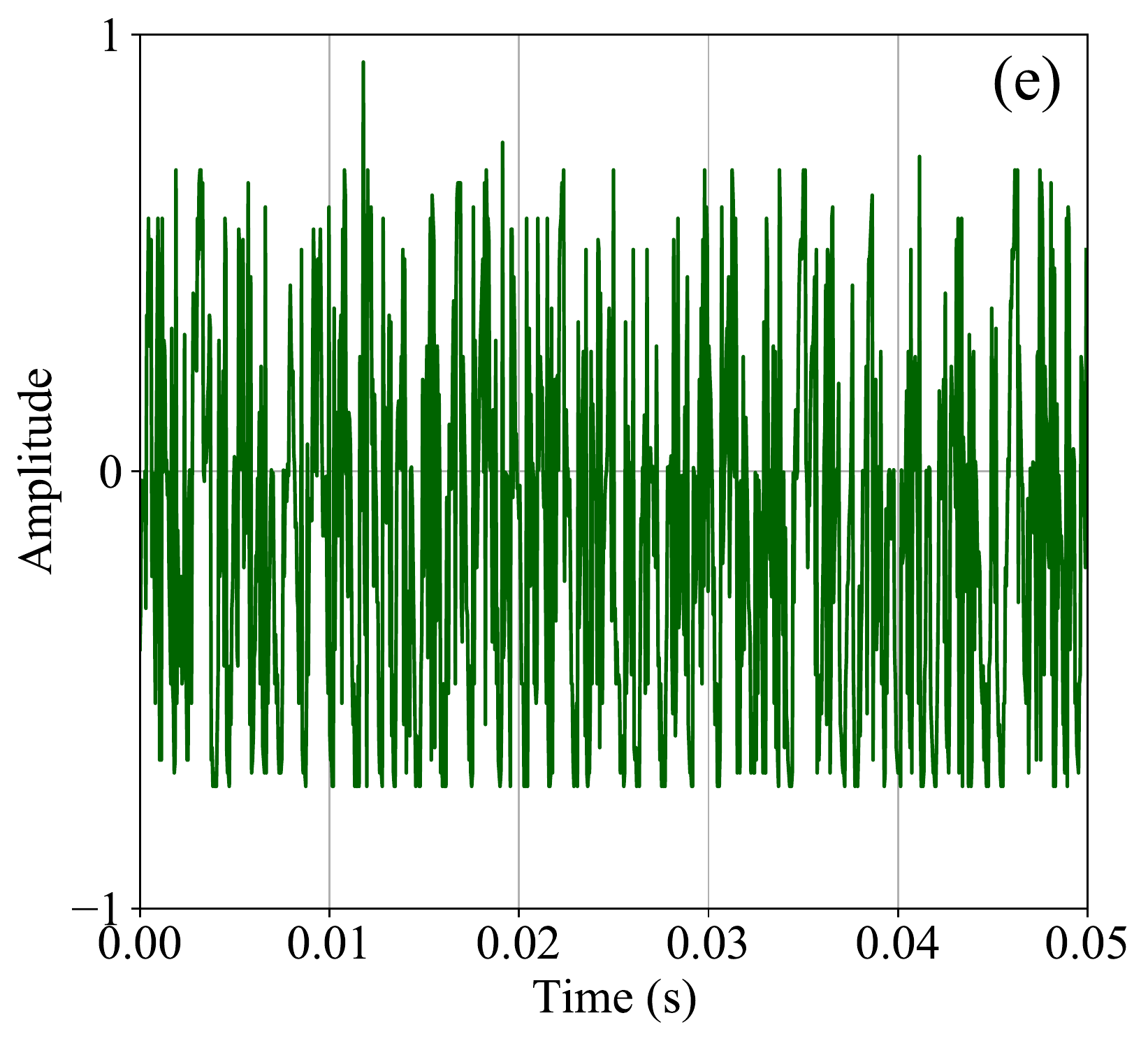}
& \includegraphics{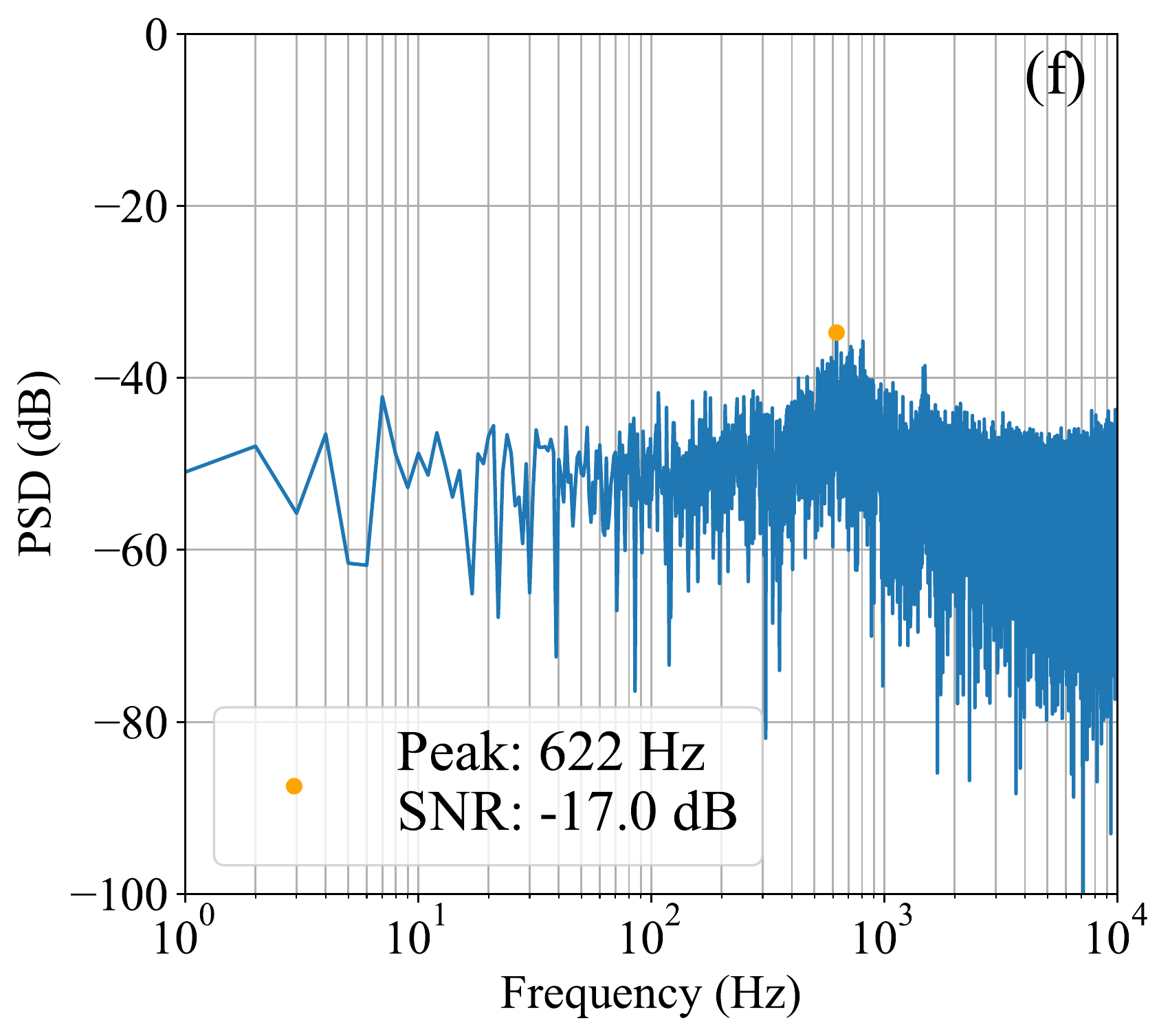} 
\end{tabular}%
}
\caption{Waveform and PSD of 500~Hz sinusoid generated by pQPNets with {\it dense factors} $2^3$ ((a), (b)), $2^0$ ((c), (d)), and $2^6$ ((e), (f)).}
\label{fig:sin_dense}
\end{figure}

\begin{figure}[t]
\resizebox{1.0\columnwidth}{!}{%
\begin{tabular}{@{}cc@{}}
  \includegraphics{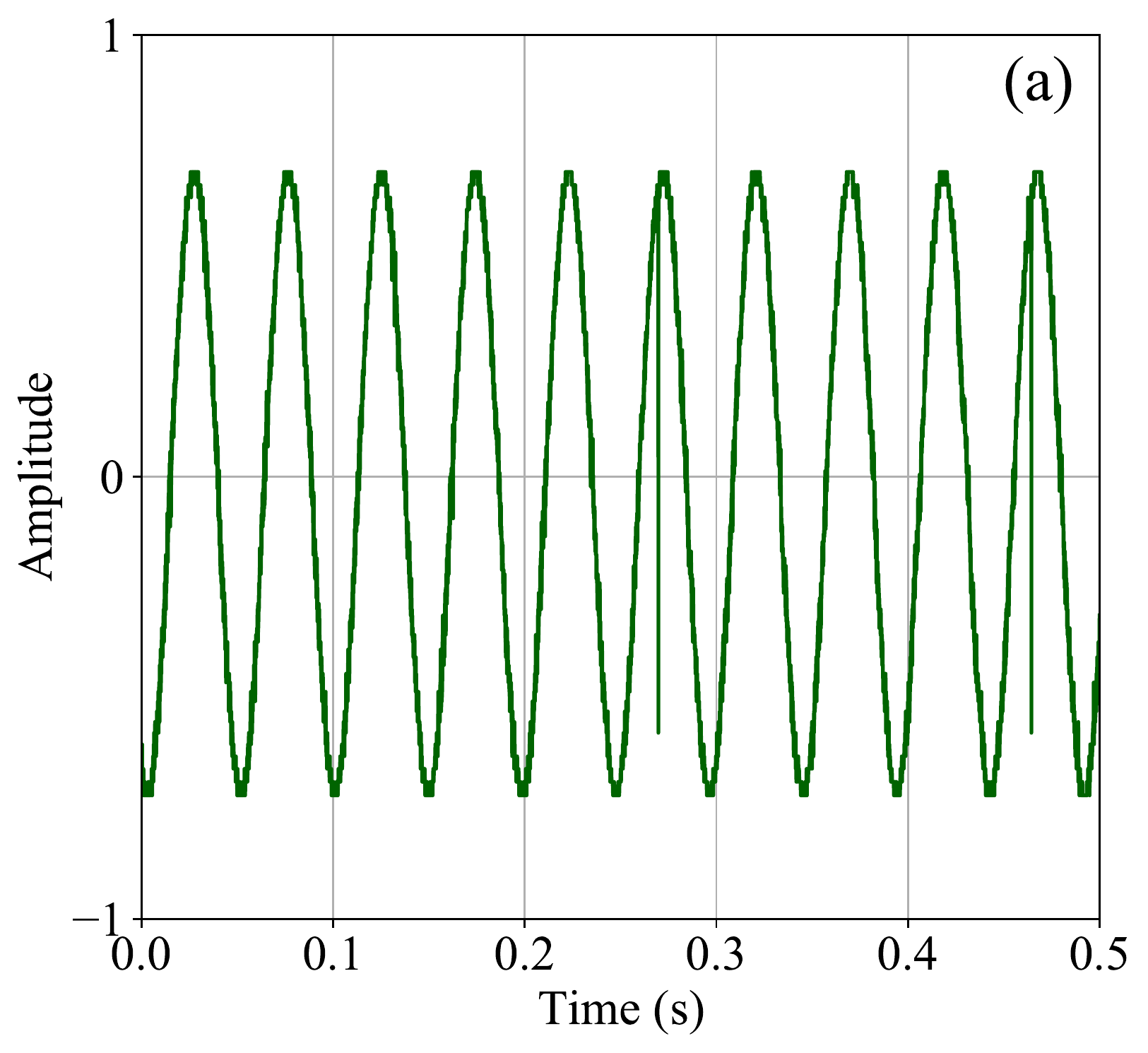}
&  \includegraphics{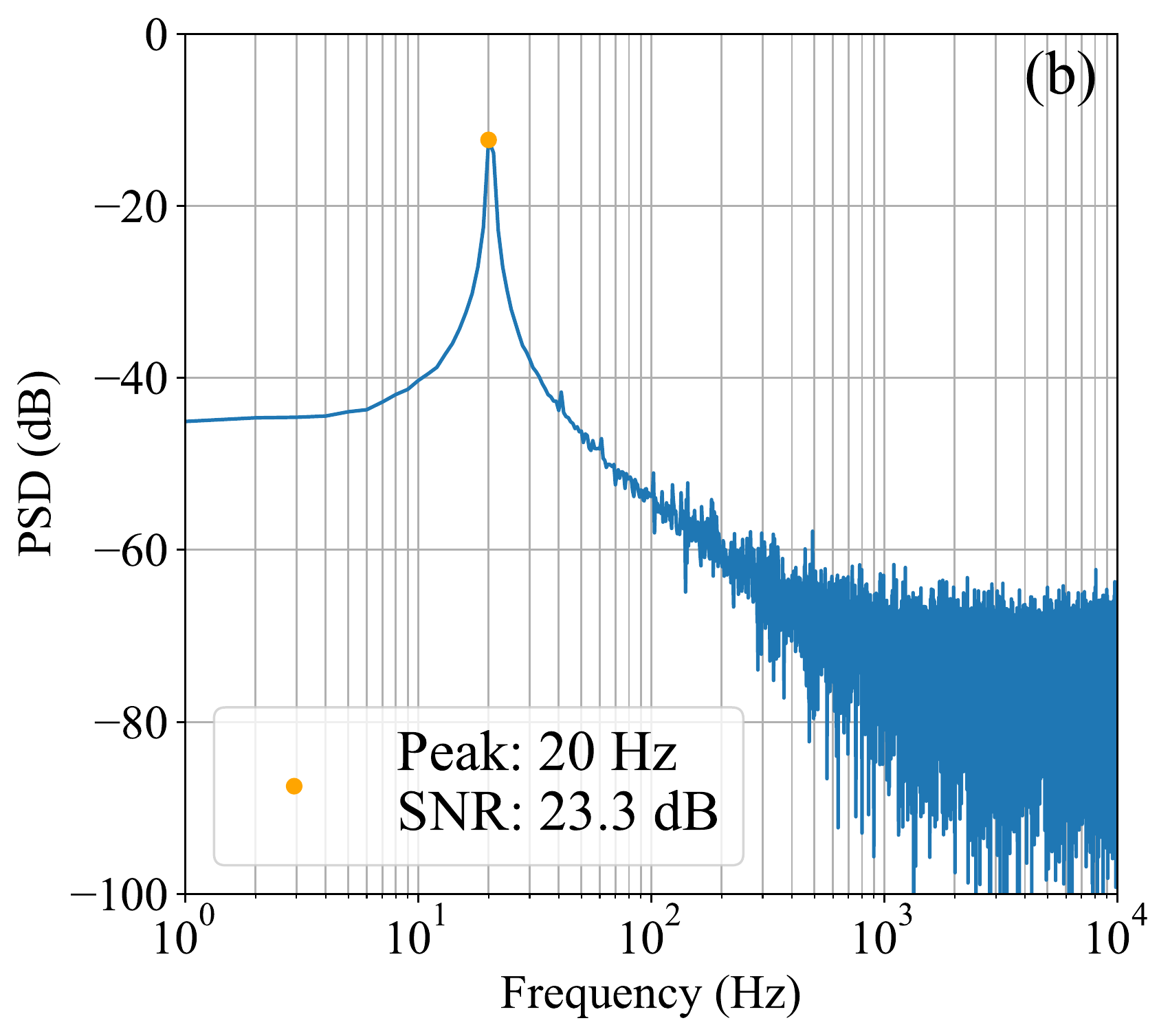} \\
  
  \includegraphics{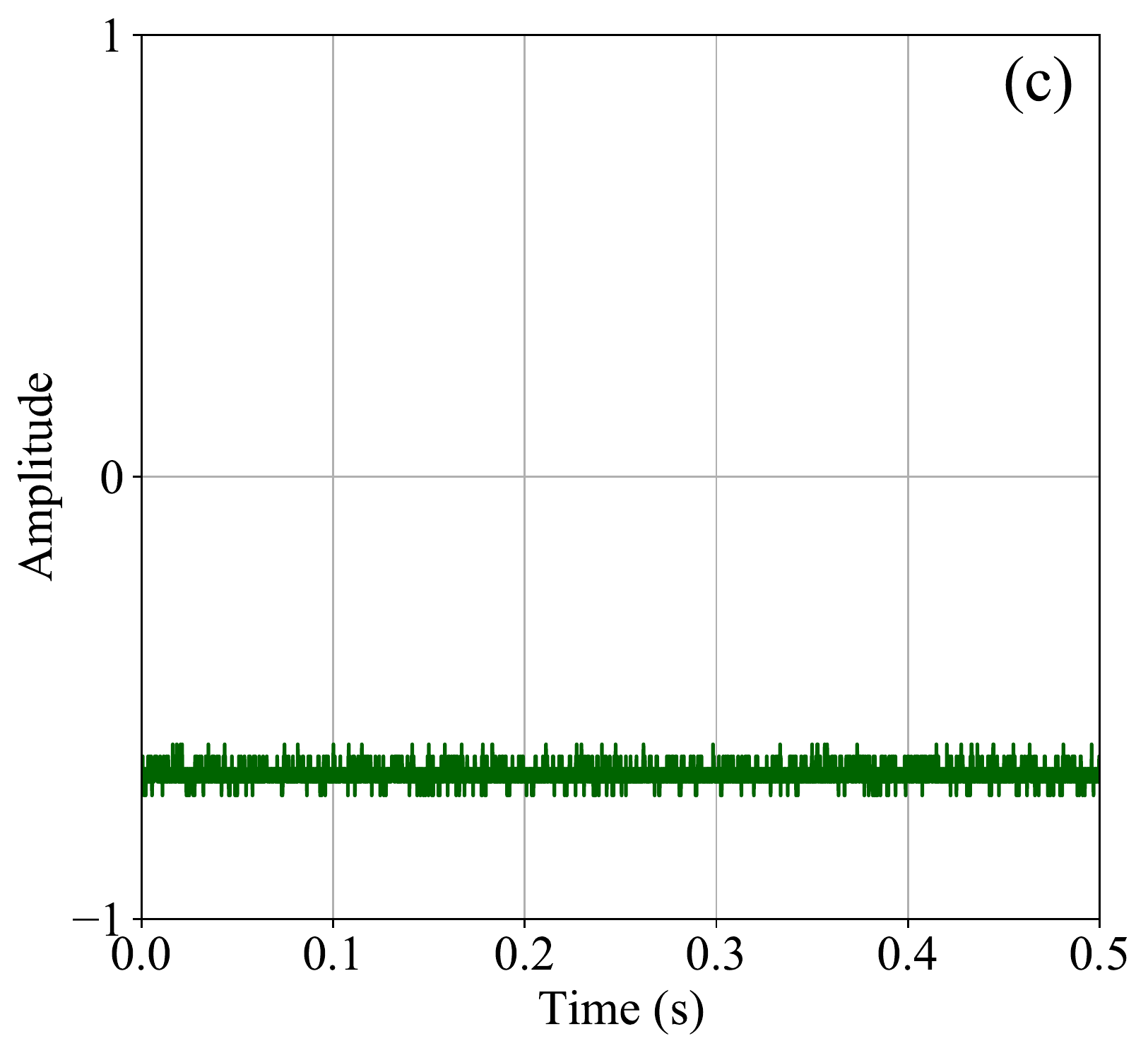}
& \includegraphics{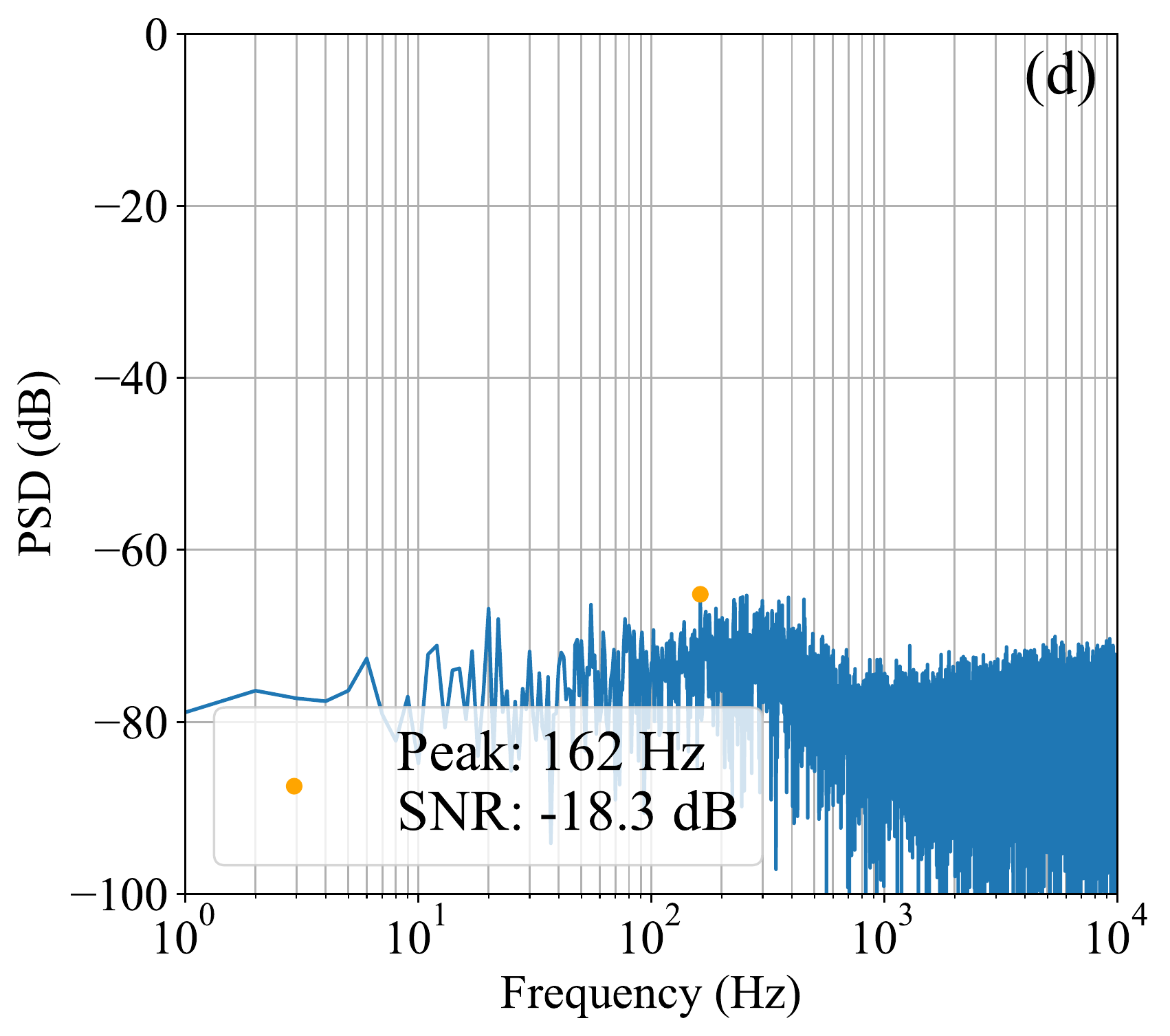} \\

  \includegraphics{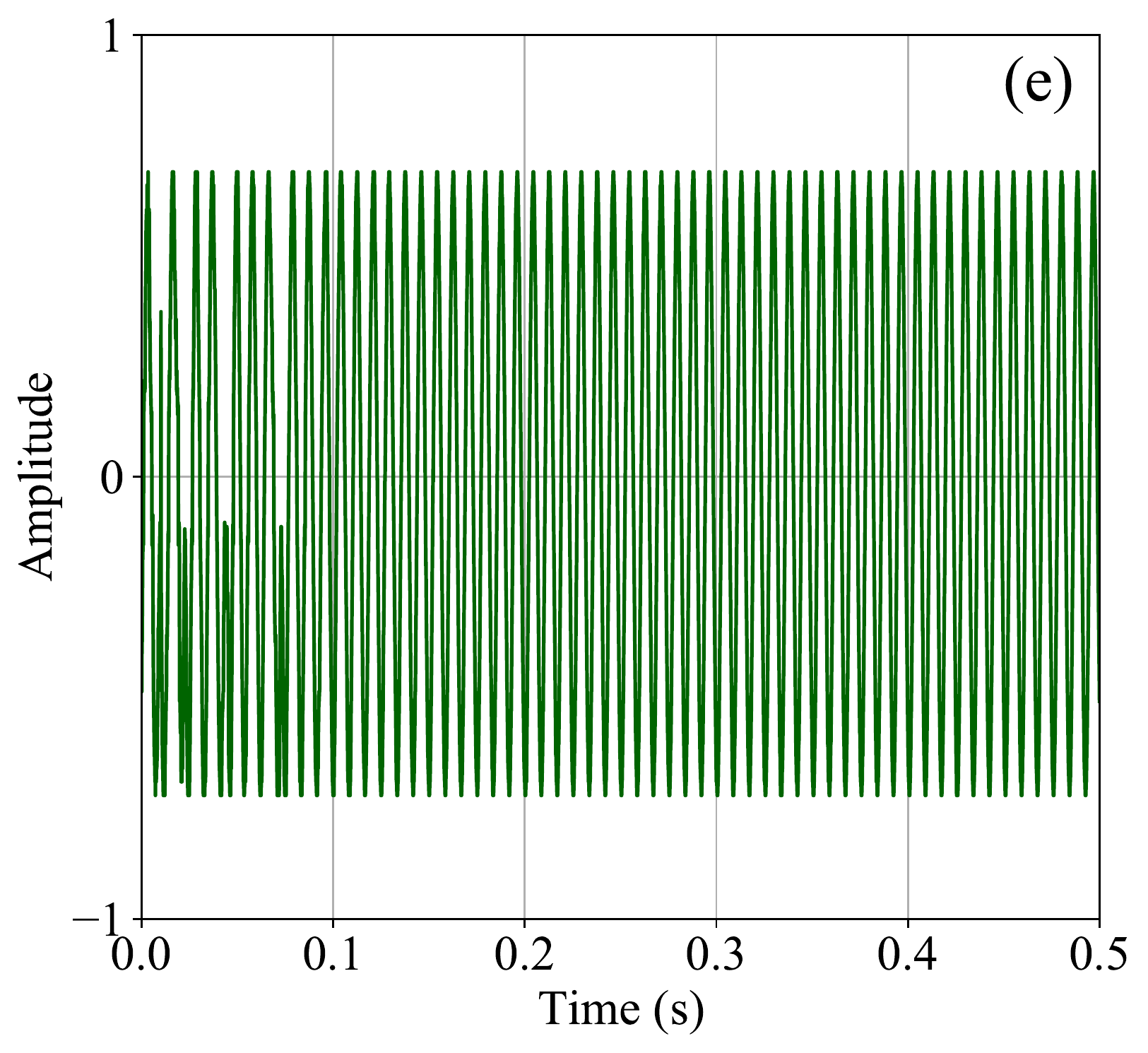}
& \includegraphics{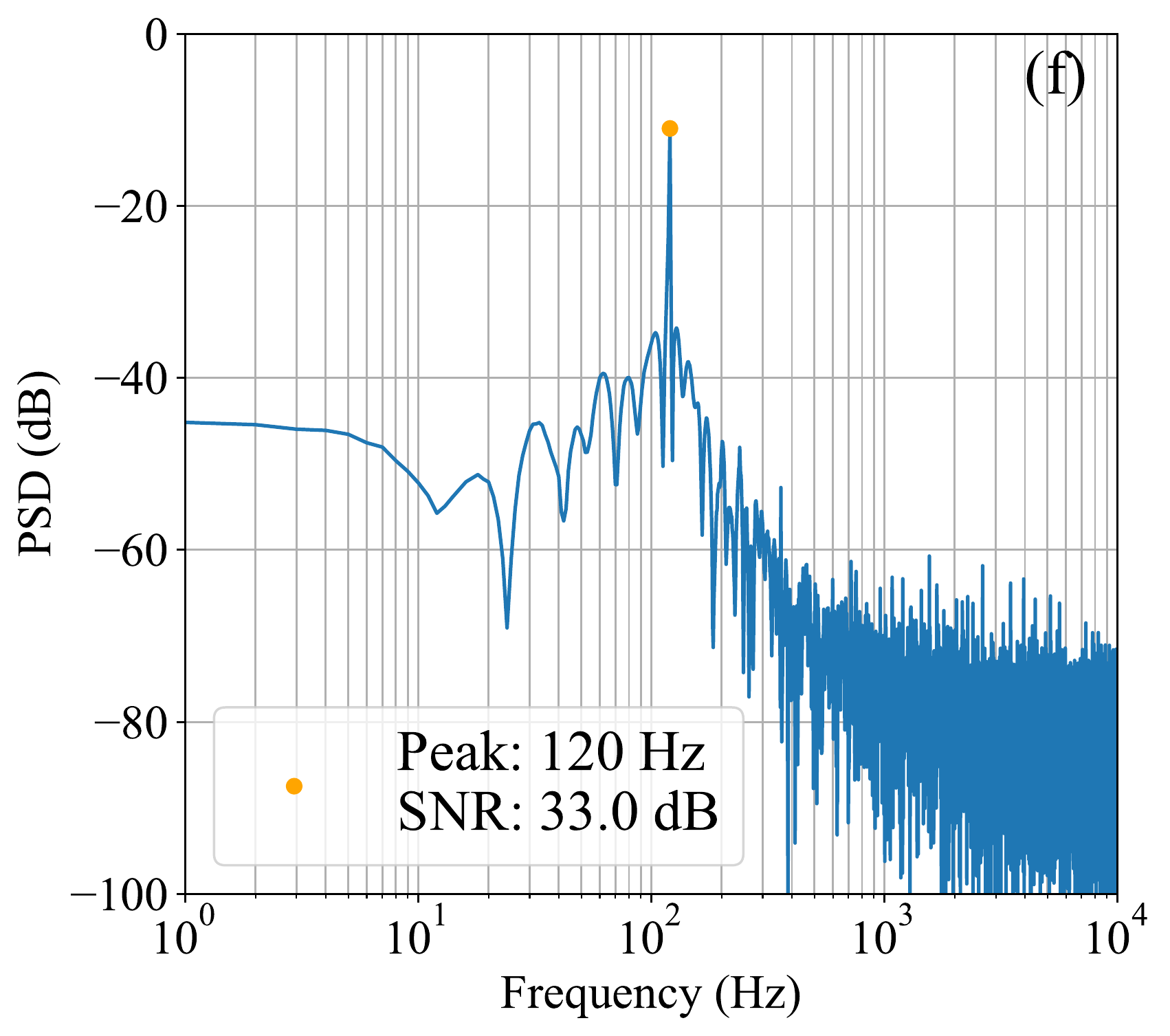} 
\end{tabular}%
}
\caption{Waveform and PSD of 20~Hz sinusoid generated by pQPNet with a {\it dense factor} $2^3$ ((a), (b)), WNc ((c), (d)), and WNf ((e), (f)).}
\label{fig:sin_model}
\end{figure}

\subsection{Discussion}
In this section, several sinusoid generation examples are presented for looking into the physical phenomena behind the objective results. As shown in Figs.~\ref{fig:sin_dense} (a) and (b), the pQPNet with a {\it dense factor} $2^3$ generated clear sine waves with an SNR 23.7~dB when conditioned on an outside auxiliary value of 500~Hz (under $3/2 U$). The PSD of this generated signal has a peak value of 502~Hz, which is very close to the ground truth, and the log $F_0$ RMSE is less than 0.01. However, the results in Figs.~\ref{fig:sin_dense} (c) and (d) show that the sine wave generated by the pQPNet with a {\it dense factor} $2^0$ includes much harmonic noise, which results in a low SNR. Even if the generated sine wave is still like a periodic signal, the wrong peak value from the second harmonic component of the PSD also causes a high log $F_0$ RMSE. Moreover, the results in Figs.~\ref{fig:sin_dense} (e) and (f) show that the pQPNet with a {\it dense factor} $2^6$ generated a very noisy signal, which results in a low SNR and an incorrect peak value of its PSD. 

In addition, as shown in Figs.~\ref{fig:sin_model} (a) and (b), the pQPNet with a {\it dense factor} $2^3$ still generated a clear sine wave with an SNR 23.3~dB and a correct peak value of its PSD when conditioned on an outside 20~Hz (under $1/2 L$) auxiliary value. However, the same-size WNc could not generate any meaningful signal, and the SNR of its generated signal is very low as shown in Figs.~\ref{fig:sin_model} (c) and (d). By contrast, the WNf still generated a clear sine wave with an SNR 33~dB but its frequency is incorrect as shown in Figs.~\ref{fig:sin_model} (e) and (f). Specifically, the PSD peak value is 120~Hz, and it implies that the WNf tends to generate seen signals even if conditioned on an unseen auxiliary feature.

The results confirm our assumptions that the high SNR and RMSE signal like Fig.~\ref{fig:sin_model} (e) is a clear sinusoid with an inaccurate frequency, the low SNR and high RMSE signal like Fig.~\ref{fig:sin_dense} (c) is a noisy sinusoid with much harmonic noise, and the very low SNR signal like Figs.~\ref{fig:sin_dense} (e) or~\ref{fig:sin_model} (c) is a noise-like signal. More results of different frequencies can be found on our demo page~\cite{demo}.

\section{Speech Generation Evaluations} \label{test_2}

In this section, we evaluate the effectiveness of the PDCNNs for speech generation. The appropriate proportions of adaptive and fixed residual blocks, the continuous pitch-dependent dilated factor, and the order of the macroblocks are explored.

\subsection{Model Architecture}
The quality of speech generation was evaluated on the basis of 11 vocoders, which included three types of vocoder, WN, QPNet, and WORLD. First, to explore the efficient {\it receptive field} extension by the PDCNNs, the compact-size QPNet vocoders were compared with the same-size WNc and double-size WNf vocoders. Secondly, the evaluations included eight variants of QPNet such as the models with different types of pitch-dependent dilated factor $E_t$ and the different order of the fixed and adaptive macroblocks. Specifically, the QPNet (fixed-to-adaptive macroblocks) and rQPNet (reversed adaptive-to-fixed macroblocks) vocoders with the continuous and discrete $E_t$ sequences were evaluated. For the unvoiced frames, the discrete $E_t$ sequence was set to ones, and the continuous $E_t$  sequence was calculated using interpolated $F_0$ values as mentioned in Section~\ref{qpnet}. Moreover, the full-size QPNet and rQPNet vocoders, which were full-size WN vocoders cascaded with four extra adaptive residual blocks, were also taken into consideration to explore the effect of the ratio of adaptive to fixed residual blocks. Last, the conventional vocoder WORLD was also adopted as a reference.

\begin{table}[t]
\fontsize{9pt}{10.8pt}
\selectfont
\caption{Architecture of Speech Generative Model}
\label{tb:model2}
{%
\begin{tabularx}{\columnwidth}{@{}p{2.5cm}YYYY@{}}
\toprule
                & WNf & WNc & (r)QPNet 
                & \begin{tabular}[c]{@{}c@{}}Full-size\\
                (r)QPNet\end{tabular} \\ \midrule
Fixed chunk     & 3  & 4  & 3  & 3  \\
Fixed block     & 10 & 4  & 4  & 10 \\
Adaptive chunk  & -  & -  & 1  & 1  \\
Adaptive block  & -  & -  & 4  & 4  \\
CNN$^1$ channel   & \multicolumn{4}{c}{512} \\
CNN$^2$ channel   & \multicolumn{4}{c}{512} \\
CNN$^3$ channel   & \multicolumn{4}{c}{256} \\
Size ($\times 10^6$) & 44 & 24 & 24 & 50 \\ \bottomrule
\end{tabularx}%
}
\\$^1$Causal and dilated CNN 
\\$^2 1 \times 1$ CNN in residual block
\\$^3 1 \times 1$ CNN in output layer
\end{table}

The network architectures and model sizes are shown in Table~\ref{tb:model2}. The learning rate was $1\times 10^{-4}$ without decay, the minibatch size was one, the batch length was 20,000 samples, and the optimizer was Adam~\cite{adam} for all models. Since even the compact-size WNc had tens of millions of parameters, which was the same order of magnitude as that of WNf, the training iterations were empirically set to 200,000 for all models. Note that we did not evaluate speech generation using the pQPNet model because it failed to model the short-term correlation of speech according to our internal experiments.

\subsection{Evaluation Setting}
All models were trained in a multispeaker manner. The training corpus of these multispeaker NN-based vocoders consisted of the training sets of the "bdl" and "slt" speakers of CMU-ARCTIC~\cite{arctic} and all speakers of VCC2018~\cite{vcc2018}. The total number of training utterances was around 3000, and the total training data length was around three hours. The evaluation corpus was composed of the SPOKE set of VCC2018, which included two female and two male speakers, and each speaker had 35 test utterances. All speech data were set to a sampling rate of 22,050~Hz and a 16-bit resolution. The waveform signals for the categorical output of the NN-based vocoders were further encoded into 8~bits using the $\mu$-law. The 513-dimensional spectral ($sp$) and $ap$ and one-dimensional $F_0$ features were extracted using WORLD. The $sp$ feature was further parameterized into 34-dimensional $mcep$, $ap$ was coded into two-dimensional components, and $F_0$ was converted into continuous $F_0$ and the voice/unvoice ($U/V$) binary code for the auxiliary features~\cite{sd_wn_vocoder}. The $F_0$ range of the SPOKE set was around 40--330~Hz, and the $F_0$ mean was around 150~Hz. The unseen outside auxiliary features were simulated by replacing the original $F_0$ values of the acoustic features with the scaled $F_0$ values, and the scaling ratios were 1/2, 3/4, 5/4, 3/2, and 2. A demo can be found on our demo page~\cite{demo}, and the open-source QPNet implementation\footnote{https://github.com/bigpon/QPNet} has been released.

\begin{table}[t]
\fontsize{9pt}{10.8pt}
\selectfont
\caption{QPNet with Different Dense Factors}
\label{tb:dense_obj}
{%
\begin{tabularx}{\columnwidth}{@{}p{1.4cm}YYYYYYY@{}}
\toprule
Dense $a$      & $2^0$ & $2^1$ & $2^2$ & $2^3$ & $2^4$ & $2^5$ & $2^6$   \\ \midrule
MCD (dB)       & 4.05  & \textbf{4.02}  & 4.03  & 4.08  & 4.17  & 4.63  & 4.26  \\
$F_0$RMSE      & 0.23  & 0.17  & 0.15  & \textbf{0.13}  & 0.14  & 0.21  & 0.24  \\
$U/V$ (\%)     & 21.8  & 16.0  & 14.2  & \textbf{13.2}  & 13.5  & 20.9  & 19.3  \\ \bottomrule
\end{tabularx}%
}
\end{table}

\begin{table}[t]
\fontsize{9pt}{10.8pt}
\selectfont
\caption{Effective Receptive Field Length (samples)}
\label{tb:dense_erf}
{%
\begin{tabularx}{\columnwidth}{@{}p{1.1cm}YYYYYYY@{}}
\toprule
Dense $a$      & $2^0$ & $2^1$ & $2^2$ & $2^3$ & $2^4$ & $2^5$ & $2^6$   \\ \midrule
Length         & \textbf{2753} & 1399 & 723  & 384  & 215  & 130  & 88   \\
               & $\pm$~8.3 & $\pm$~4.2 & $\pm$~2.1 & $\pm$~1.0 
               & $\pm$~0.5 & $\pm$~0.3 & $\pm$~0.1 \\ \bottomrule
\end{tabularx}%
}
\end{table}

\subsection{Objective Evaluation}
For the objective evaluations, the ground truth acoustic features were extracted from natural speech utterances using WORLD, and the extraction error from WORLD was neglected. A speaker-dependent $F_0$ range was applied to the feature extraction of each speaker to improve the extraction accuracy, and the $F_0$ range was set following the process in~\cite{sprocket}. Since WORLD was developed to extract $F_0$-independent spectral features~\cite{world}, the WORLD-extracted $sp$ feature was assumed to be highly uncorrelated to the $F_0$ feature in this paper. Therefore, the ground truth acoustic features for the scaled $F_0$ scenarios were the same natural spectral features with the $F_0$ feature scaled by an assigned ratio. The auxiliary features of the evaluated vocoders were the ground truth acoustic features. Mel-cepstral distortion (MCD) was applied to measure the spectral reconstruction capability of the vocoders, and the MCD was calculated between the auxiliary $mcep$ and the WORLD-extracted $mcep$ from the generated speech. The pitch accuracy of the generated speech was evaluated using the RMSE of the auxiliary $F_0$ and the WORLD-extracted $F_0$ value from the generated speech in the logarithmic domain. The unvoiced/voiced ($U/V$) decision error was also taken into account in the evaluation of the prosodic prediction capability, which was the percentage of the unvoiced/voiced decision difference of each utterance.

\begin{table*}[t]
\fontsize{9pt}{10.8pt}
\selectfont
\caption{MCD ({\rm dB}) with Frame-based 95~\% Confidence Interval (CI) of Different Generation Models for Speech Generation}
\label{tb:mcd}
{%
\begin{tabularx}{\textwidth}{@{}p{1.5cm}YYYYYYYYYYY@{}}
\toprule
 &
  \multicolumn{1}{c}{WORLD} &
  \multicolumn{1}{c}{WNc} &
  \multicolumn{1}{c}{WNf} &
  \multicolumn{2}{c}{QPNet} &
  \multicolumn{2}{c}{Full-size QPNet} &
  \multicolumn{2}{c}{rQPNet} &
  \multicolumn{2}{c}{Full-size rQPNet} \\ \midrule
$E_t$ &
  \multicolumn{1}{c}{-} &
  \multicolumn{1}{c}{-} &
  \multicolumn{1}{c}{-} &
  \multicolumn{1}{c}{cont.} &
  \multicolumn{1}{c}{disc.} &
  \multicolumn{1}{c}{cont.} &
  \multicolumn{1}{c}{disc.} &
  \multicolumn{1}{c}{cont.} &
  \multicolumn{1}{c}{disc.} &
  \multicolumn{1}{c}{cont.} &
  \multicolumn{1}{c}{disc.} \\ \midrule
\multirow{2}{*}{$1\times F_0$}
& \cellcolor[HTML]{EFEFEF}2.51 & 4.34 & 3.58 & 4.08 
& 4.16 & 3.59 & 3.60 & 3.91 
& 3.97 & \textbf{3.54} & 3.58 \\
& \cellcolor[HTML]{EFEFEF}$\pm$~0.009 & $\pm$~0.008 & $\pm$~0.007 & $\pm$~0.007 
& $\pm$~0.007 & $\pm$~0.008 & $\pm$~0.007 & $\pm$~0.007 
& $\pm$~0.008 & $\pm$~0.007 & $\pm$~0.007 \\
\multirow{2}{*}{$1/2\times F_0$}  
& \cellcolor[HTML]{EFEFEF}3.88 & 5.02 & 4.56 & 4.79 
& 4.90 & 4.49 & 4.46 & 4.66
& 4.79 & 4.43 & \textbf{4.40} \\
& \cellcolor[HTML]{EFEFEF}$\pm$~0.016 & $\pm$~0.009 & $\pm$~0.008 & $\pm$~0.008
& $\pm$~0.008 & $\pm$~0.008 & $\pm$~0.008 & $\pm$~0.008
& $\pm$~0.009 & $\pm$~0.008 & $\pm$~0.008 \\
\multirow{2}{*}{$3/4\times F_0$}  
& \cellcolor[HTML]{EFEFEF}2.91 & 4.58 & 3.95 & 4.34
& 4.43 & 3.95 & 3.91 & 4.19
& 4.26 & \textbf{3.87} & 3.88 \\
& \cellcolor[HTML]{EFEFEF}$\pm$~0.012 & $\pm$~0.009 & $\pm$~0.008 & $\pm$~0.008
& $\pm$~0.008 & $\pm$~0.008 & $\pm$~0.008 & $\pm$~0.008
& $\pm$~0.008 & $\pm$~0.008 & $\pm$~0.008 \\
\multirow{2}{*}{$5/4\times F_0$}  
& \cellcolor[HTML]{EFEFEF}2.76 & 4.39 & 3.62 & 4.16
& 4.25 & \textbf{3.54} & 3.60 & 3.98
& 4.03 & 3.60 & 3.63 \\
& \cellcolor[HTML]{EFEFEF}$\pm$~0.008 & $\pm$~0.009 & $\pm$~0.007 & $\pm$~0.007
& $\pm$~0.007 & $\pm$~0.008 & $\pm$~0.007 & $\pm$~0.007
& $\pm$~0.008 & $\pm$~0.007 & $\pm$~0.007 \\
\multirow{2}{*}{$3/2\times F_0$}  
& \cellcolor[HTML]{EFEFEF}3.04 & 4.50 & 3.68 & 4.27
& 4.35 & \textbf{3.56} & 3.64 & 4.06
& 4.12 & 3.65 & 3.67 \\
& \cellcolor[HTML]{EFEFEF}$\pm$~0.009 & $\pm$~0.009 & $\pm$~0.007 & $\pm$~0.007
& $\pm$~0.008 & $\pm$~0.007 & $\pm$~0.007 & $\pm$~0.007
& $\pm$~0.008 & $\pm$~0.007 & $\pm$~0.007 \\
\multirow{2}{*}{$2\times F_0$}    
& \cellcolor[HTML]{EFEFEF}3.75 & 4.75 & 3.86 & 4.59
& 4.64 & \textbf{3.82} & 3.88 & 4.33
& 4.37 & 3.92 & 3.90 \\
& \cellcolor[HTML]{EFEFEF}$\pm$~0.010 & $\pm$~0.008 & $\pm$~0.007 & $\pm$~0.008
& $\pm$~0.008 & $\pm$~0.007 & $\pm$~0.007 & $\pm$~0.008
& $\pm$~0.008 & $\pm$~0.007 & $\pm$~0.007 \\ \midrule
\multirow{2}{*}{Average} 
& \cellcolor[HTML]{EFEFEF}3.14 & 4.60 & 3.87 & 4.37
& 4.45 & \textbf{3.83} & 3.85 & 4.19
& 4.26 & 3.84 & 3.84 \\
& \cellcolor[HTML]{EFEFEF}$\pm$~0.005 & $\pm$~0.004 & $\pm$~0.003 & $\pm$~0.003
& $\pm$~0.003 & $\pm$~0.003 & $\pm$~0.003 & $\pm$~0.003
& $\pm$~0.004 & $\pm$~0.003 & $\pm$~0.003 \\ \bottomrule
\end{tabularx}%
}
\end{table*}

\begin{table*}[t]
\caption{Log $F_0$ RMSE with Utterance-based 95~\% CI of Different Generation Models for Speech Generation}
\label{tb:rmse}
\fontsize{9pt}{10.8pt}
\selectfont
{%
\begin{tabularx}{\textwidth}{@{}p{1.5cm}YYYYYYYYYYY@{}}
\toprule
 &
  \multicolumn{1}{c}{WORLD} &
  \multicolumn{1}{c}{WNc} &
  \multicolumn{1}{c}{WNf} &
  \multicolumn{2}{c}{QPNet} &
  \multicolumn{2}{c}{Full-size QPNet} &
  \multicolumn{2}{c}{rQPNet} &
  \multicolumn{2}{c}{Full-size rQPNet} \\ \midrule
$E_t$ &
  \multicolumn{1}{c}{-} &
  \multicolumn{1}{c}{-} &
  \multicolumn{1}{c}{-} &
  \multicolumn{1}{c}{cont.} &
  \multicolumn{1}{c}{disc.} &
  \multicolumn{1}{c}{cont.} &
  \multicolumn{1}{c}{disc.} &
  \multicolumn{1}{c}{cont.} &
  \multicolumn{1}{c}{disc.} &
  \multicolumn{1}{c}{cont.} &
  \multicolumn{1}{c}{disc.} \\ \midrule
\multirow{2}{*}{$1\times F_0$}
& \cellcolor[HTML]{EFEFEF}0.09  & 0.26  & 0.14  & \textbf{0.13}
& 0.14  & 0.15  & 0.15  & 0.16
& 0.16  & 0.15  & 0.15  \\
& \cellcolor[HTML]{EFEFEF}$\pm$~0.007 & $\pm$~0.026 & $\pm$~0.011 & $\pm$~0.010 
& $\pm$~0.011 & $\pm$~0.014 & $\pm$~0.014 & $\pm$~0.016 
& $\pm$~0.015 & $\pm$~0.013 & $\pm$~0.011 \\
\multirow{2}{*}{$1/2\times F_0$}
& \cellcolor[HTML]{EFEFEF}0.13  & 0.38  & 0.30  & \textbf{0.23}
& 0.24  & 0.33  & 0.34  & 0.26
& 0.26  & 0.33  & 0.33  \\
& \cellcolor[HTML]{EFEFEF}$\pm$~0.013 & $\pm$~0.026 & $\pm$~0.029 & $\pm$~0.024 
& $\pm$~0.026 & $\pm$~0.035 & $\pm$~0.036 & $\pm$~0.027 
& $\pm$~0.027 & $\pm$~0.034 & $\pm$~0.034 \\
\multirow{2}{*}{$3/4\times F_0$}
& \cellcolor[HTML]{EFEFEF}0.10  & 0.32  & 0.20  & \textbf{0.17}
& 0.18  & 0.22  & 0.22  & 0.21
& 0.22  & 0.21  & 0.21  \\
& \cellcolor[HTML]{EFEFEF}$\pm$~0.009 & $\pm$~0.026 & $\pm$~0.016 & $\pm$~0.015 
& $\pm$~0.018 & $\pm$~0.021 & $\pm$~0.020 & $\pm$~0.021 
& $\pm$~0.022 & $\pm$~0.017 & $\pm$~0.017 \\
\multirow{2}{*}{$5/4\times F_0$}
& \cellcolor[HTML]{EFEFEF}0.09  & 0.25  & 0.17  & 0.14
& \textbf{0.13}  & 0.15  & 0.15  & 0.16
& 0.16  & 0.15  & 0.16  \\
& \cellcolor[HTML]{EFEFEF}$\pm$~0.008 & $\pm$~0.017 & $\pm$~0.009 & $\pm$~0.009 
& $\pm$~0.008 & $\pm$~0.010 & $\pm$~0.010 & $\pm$~0.011 
& $\pm$~0.010 & $\pm$~0.010 & $\pm$~0.009 \\
\multirow{2}{*}{$3/2\times F_0$}
& \cellcolor[HTML]{EFEFEF}0.09  & 0.27  & 0.21  & 0.16
& \textbf{0.15}  & 0.19  & 0.19  & 0.18
& 0.19  & 0.20  & 0.20  \\
& \cellcolor[HTML]{EFEFEF}$\pm$~0.008 & $\pm$~0.014 & $\pm$~0.008 & $\pm$~0.009 
& $\pm$~0.008 & $\pm$~0.009 & $\pm$~0.010 & $\pm$~0.009 
& $\pm$~0.009 & $\pm$~0.011 & $\pm$~0.010 \\
\multirow{2}{*}{$2\times F_0$}
& \cellcolor[HTML]{EFEFEF}0.09  & 0.28  & 0.26  & 0.18
& \textbf{0.17}  & 0.26  & 0.26  & 0.18
& 0.20  & 0.29  & 0.33  \\
& \cellcolor[HTML]{EFEFEF}$\pm$~0.008 & $\pm$~0.013 & $\pm$~0.014 & $\pm$~0.024 
& $\pm$~0.008 & $\pm$~0.015 & $\pm$~0.039 & $\pm$~0.008 
& $\pm$~0.010 & $\pm$~0.048 & $\pm$~0.050 \\ \midrule
\multirow{2}{*}{Average}
& \cellcolor[HTML]{EFEFEF}0.10  & 0.29  & 0.21  & \textbf{0.17}
& \textbf{0.17}  & 0.22  & 0.22  & 0.19
& 0.20  & 0.22  & 0.23  \\
& \cellcolor[HTML]{EFEFEF}$\pm$~0.004 & $\pm$~0.009 & $\pm$~0.007 & $\pm$~0.007 
& $\pm$~0.006 & $\pm$~0.009 & $\pm$~0.011 & $\pm$~0.007 
& $\pm$~0.007 & $\pm$~0.011 & $\pm$~0.012 \\ \bottomrule
\end{tabularx}%
}
\end{table*}

\subsubsection{Dense Factor}
Since speech generation is more complicated than sine wave generation, we first conducted an objective evaluation of the QPNet models with different {\it dense factors} for speech generation to check the consistency of the efficient {\it dense factor} value. As shown in Table~\ref{tb:dense_obj}, the tendency of the objective evaluation is similar to the results of the sinusoid generation evaluation. That is, the QPNets with {\it dense factors} from $2^1$--$2^4$ achieved similar generative performance while the speech quality and pitch accuracy of the QPNets with {\it dense factors} $2^5$ and $2^6$ markedly degraded because of the much shorter {\it effective receptive field} lengths. Specifically, as shown in Table~\ref{tb:dense_erf}, the average {\it effective receptive field} lengths of the QPNets with the {\it dense factors} $2^5$ and $2^6$ are much shorter than others, and the lengths were too short to cover at least one cycle of the signal with 150~Hz, which was the $F_0$ mean of the SPOKE set. 

Furthermore, although the QPNet with a $2^0$ {\it dense factor} had the longest average {\it effective receptive field} length and achieved an acceptable MCD, the higher RMSE of log $F_0$ and $U/V$ error indicate its instability, which was also observed in the sinusoid generation evaluation. The results also confirm our assumption that the QPNet with a $2^0$ {\it dense factor} cannot model the periodic components well because the Nyquist frequency of the QPNet adaptive macroblock is lower than the bandwidth of the periodic components. Moreover, because of the natural fluctuations of speech, $F_0$ extraction errors, etc., the oversampling models with an appropriate {\it dense factors} such as $2^2$--$2^4$, which keep long enough {\it effective receptive fields},  also achieve better performance. As a result, the {\it dense factors} of the following QPNet-series models were set to $2^3$ because of the lowest RMSE of log $F_0$ and $U/V$ error with an acceptable MCD. The internal subjective evaluation results also show the preference of the utterances generated by the QPNet with the {\it dense factor} $2^3$.

\begin{table*}[t]
\caption{$U/V$ Decision Error Rate (\%) with Utterance-based 95~\% CI of Different Generation Models for Speech Generation}
\label{tb:uv}
\fontsize{9pt}{10.8pt}
\selectfont
{%
\begin{tabularx}{\textwidth}{@{}p{1.5cm}YYYYYYYYYYY@{}}
\toprule
 &
  \multicolumn{1}{c}{WORLD} &
  \multicolumn{1}{c}{WNc} &
  \multicolumn{1}{c}{WNf} &
  \multicolumn{2}{c}{QPNet} &
  \multicolumn{2}{c}{Full-size QPNet} &
  \multicolumn{2}{c}{rQPNet} &
  \multicolumn{2}{c}{Full-size rQPNet} \\ \midrule
$E_t$ &
  \multicolumn{1}{c}{-} &
  \multicolumn{1}{c}{-} &
  \multicolumn{1}{c}{-} &
  \multicolumn{1}{c}{cont.} &
  \multicolumn{1}{c}{disc.} &
  \multicolumn{1}{c}{cont.} &
  \multicolumn{1}{c}{disc.} &
  \multicolumn{1}{c}{cont.} &
  \multicolumn{1}{c}{disc.} &
  \multicolumn{1}{c}{cont.} &
  \multicolumn{1}{c}{disc.} \\ \midrule
\multirow{2}{*}{$1\times F_0$}
& \cellcolor[HTML]{EFEFEF}9.9  & 23.6 & 14.5 & \textbf{13.2} & 13.9 & 14.9
& 14.3 & 15.7 & 15.2 & 14.0 & 14.7 \\
& \cellcolor[HTML]{EFEFEF}$\pm$~0.79 & $\pm$~1.86 & $\pm$~1.04 & $\pm$~0.96 & $\pm$~1.07 & $\pm$~1.12
& $\pm$~1.08 & $\pm$~1.21 & $\pm$~1.10 & $\pm$~1.05 & $\pm$~1.10 \\
\multirow{2}{*}{$1/2\times F_0$}
& \cellcolor[HTML]{EFEFEF}16.0 & 35.0 & 26.6 & \textbf{22.3} & 22.8 & 29.9
& 30.1 & 27.6 & 26.3 & 29.5 & 30.4 \\
& \cellcolor[HTML]{EFEFEF}$\pm$~1.04 & $\pm$~1.35 & $\pm$~1.49 & $\pm$~1.34 & $\pm$~1.40 & $\pm$~1.78
& $\pm$~1.52 & $\pm$~1.50 & $\pm$~1.22 & $\pm$~1.38 & $\pm$~1.63 \\
\multirow{2}{*}{$3/4\times F_0$}
& \cellcolor[HTML]{EFEFEF}12.2 & 29.1 & 18.2 & \textbf{16.4} & 17.5 & 20.2
& 20.2 & 19.8 & 20.2 & 18.5 & 19.5 \\
& \cellcolor[HTML]{EFEFEF}$\pm$~0.92 & $\pm$~1.56 & $\pm$~1.39 & $\pm$~1.22 & $\pm$~1.28 & $\pm$~1.45 
& $\pm$~1.43 & $\pm$~1.65 & $\pm$~1.51 & $\pm$~1.14 & $\pm$~1.30 \\
\multirow{2}{*}{$5/4\times F_0$}
& \cellcolor[HTML]{EFEFEF}9.6  & 24.9 & 13.3 & \textbf{13.1} & 13.9 & 13.9
& 13.5 & 14.5 & 14.2 & 14.1 & 13.9 \\
& \cellcolor[HTML]{EFEFEF}$\pm$~0.63 & $\pm$~1.92 & $\pm$~0.91 & $\pm$~0.99 & $\pm$~1.01 & $\pm$~1.00 
& $\pm$~0.99 & $\pm$~1.07 & $\pm$~1.09 & $\pm$~1.04 & $\pm$~0.96 \\
\multirow{2}{*}{$3/2\times F_0$}
& \cellcolor[HTML]{EFEFEF}9.9  & 27.9 & 13.8 & 14.7 & 15.5 & 13.6
& 14.8 & 16.3 & 15.7 & \textbf{13.3} & 14.8 \\
& \cellcolor[HTML]{EFEFEF}$\pm$~0.69 & $\pm$~1.78 & $\pm$~1.04 & $\pm$~0.96 & $\pm$~1.12 & $\pm$~1.01
& $\pm$~1.22 & $\pm$~1.15 & $\pm$~1.14 & $\pm$~0.87 & $\pm$~1.01 \\
\multirow{2}{*}{$2\times F_0$}
& \cellcolor[HTML]{EFEFEF}10.5 & 36.7 & \textbf{20.3} & 21.9 & 20.6 & 26.2
& 24.3 & 25.3 & 26.3 & 29.6 & 33.4 \\
& \cellcolor[HTML]{EFEFEF}$\pm$~0.56 & $\pm$~1.80 & $\pm$~1.56 & $\pm$~1.93 & $\pm$~1.71 & $\pm$~2.32
& $\pm$~1.97 & $\pm$~2.33 & $\pm$~2.55 & $\pm$~2.87 & $\pm$~3.26 \\ \midrule
\multirow{2}{*}{Average}
& \cellcolor[HTML]{EFEFEF}11.3 & 29.5 & 17.8 & \textbf{16.9} & 17.4 & 19.8
& 19.5 & 19.8 & 19.7 & 19.8 & 21.1 \\
& \cellcolor[HTML]{EFEFEF}$\pm$~0.35 & $\pm$~0.77 & $\pm$~0.60 & $\pm$~0.58 & $\pm$~0.57 & $\pm$~0.75
& $\pm$~0.70 & $\pm$~0.71 & $\pm$~0.70 & $\pm$~0.79 & $\pm$~0.88 \\ \bottomrule
\end{tabularx}%
}
\end{table*}

\subsubsection{Spectral Accuracy}
As shown in Table~\ref{tb:mcd}, in terms of spectral prediction capability, the compact-size (r)QPNet vocoders with the proposed PDCNNs significantly outperformed the same-size WNc vocoder. The results confirm the effectiveness of the QP structure to skip some redundant samples using the prior pitch knowledge for a more efficient {\it receptive field} extension. However, the MCDs of the double-size WNf vocoder are lower than that of the compact-size (r)QPNet vocoders, and the full-size (r)QPNet vocoders with the largest network size also outperformed the WNf vocoder in terms of MCD. The results indicate that the MCD values are highly related to the network sizes, so a deeper network attains a more powerful spectral modeling capability. Furthermore, the systems with continuous pitch-dependent dilated factors achieved lower MCDs than those with discrete ones, and the result is consistent with our internal subjective evaluation for speech quality. However, the MCD differences of the rQPNet and QPNet vocoders were not reflected in the perceptual quality, and they had similar speech qualities according to the internal evaluation.

\subsubsection{Pitch Accuracy}
The log $F_0$ RMSE results in Table~\ref{tb:rmse} also show that both the compact-size QPNet and rQPNet vocoders attained markedly higher pitch accuracy than the same-size WNc vocoder, particularly when conditioned on the unseen $F_0$ with a large shift. Since the WNf vocoder usually generates seen signals even conditioned on unseen auxiliary features, the compact-size QPNet vocoder achieved higher pitch accuracies than the WNf vocoder as expected. The results indicate that the PDCNNs with the prior pitch knowledge improved the pitch controllability of these vocoders against the unseen $F_0$. However, the pitch accuracies of the full-size (r)QPNet vocoders are lower than that of the (r)QPNet vocoders. The possible reason is that the unbalanced proportion of the adaptive and fixed residual blocks impaired the pitch controllability. That is, for the full-size (r)QPNet vocoders, the number of the fixed blocks is markedly larger than the number of the adaptive blocks. Therefore, the network might be dominated by the fixed blocks, which degraded the influence from the adaptive blocks. Specifically, for the (r)QPNet vocoders with a {\it dense factor} $2^3$, the {\it receptive field} length of the fixed blocks is 46 samples (The details of the {\it receptive field} length can be found in Discussion), and the average {\it effective receptive field} length of the adaptive blocks is 384 samples as shown in Table~\ref{tb:dense_erf}. However, for the full-size (r)QPNet vocoders, the {\it receptive field} length of the fixed blocks is 3070 samples, which was much longer than the 384 samples of the extra four adaptive blocks. Therefore, the influence of the adaptive blocks might be very limited.

\subsubsection{U/V Accuracy and Summary}
As shown in Table~\ref{tb:uv}, the compact-size QPNet vocoder attained the lowest $U/V$ decision error among all NN-based vocoders, and it indicates a higher capability to capture $U/V$ information. In conclusion, the compact-size QPNet vocoder with the proposed PDCNNs and continuous pitch-dependent dilated factors attained the highest accuracy of pitch and $U/V$ information among the evaluated NN-based vocoders. Although the compact-size QPNet vocoder did not achieve the same spectral prediction capability as the WNf vocoder according to the MCD results, it is difficult to measure a perceptual quality difference only on the basis of MCD. As a result, we subjectively evaluated the compact-size QPNet (with continuous pitch-dependent dilated factors), WNc, and WNf vocoders in the next section. Moreover, although the WORLD vocoder had the best objective evaluation results, the WORLD-generated speech usually lacks naturalness and contains buzz noise, which may not be reflected in the objective measurements. Therefore, the WORLD vocoder was also evaluated in the subjective tests. 

\subsection{Subjective Evaluation}
The subjective evaluations included the Mean Opinion Score (MOS) test for speech quality and the ABX preference test for perceptual pitch accuracy. Specifically, the naturalness of each utterance in the evaluation set for the MOS test was evaluated by several listeners by assigning scores of 1--5 to each utterance; the higher the score, the greater naturalness of the utterance. The MOS evaluation set was composed of randomly selected utterances generated by the WORLD, WNf, WNc, and QPNet vocoders, and the auxiliary features with 1/2 $F_0$, 3/2 $F_0$, and unchanged $F_0$. The compact-size QPNet vocoder with the continuous dilated factors was adopted and abbreviated as QPNet in the subjective evaluations. We randomly selected 20 utterances from the 35 test utterances of each condition and each speaker to form the MOS evaluation set, so the number of utterances in the set was 960. The mean, standard deviation, longest, and shortest lengths of the selected utterances were 4~s, 1.6~s, 8~s, and 1~s, respectively. The MOS evaluation set was divided into five subsets, and each subset was evaluated by two listeners, so the total number of listeners was 10. All listeners took the test using the same devices in the same quiet room. Although the listeners were not native speakers, they had worked on speech or audio generation research.

In the ABX preference test, the listeners compared two test utterances (A and B) with one reference utterance (X) to evaluate which testing utterance had a pitch contour more consistent with that of the reference utterance. Although the natural speech with the desired scaled $F_0$ does not exist, the conventional source-filter vocoders usually attain high pitch controllability. Therefore, the WORLD-generated utterances were taken as the references. The ABX evaluation set consisted of the same generated utterances of the WNf, QPNet, and WORLD vocoders as the MOS evaluation set. The number of ABX utterance pairs was 240, and each pair was evaluated by two of the same 10 listeners as in the MOS test. Since the ABX test focus on pitch accuracy, all listeners were asked to focus on the pitch differences and ignore the quality differences.

\begin{figure}[t]
\centering
\centerline{\includegraphics[width=0.92\columnwidth]{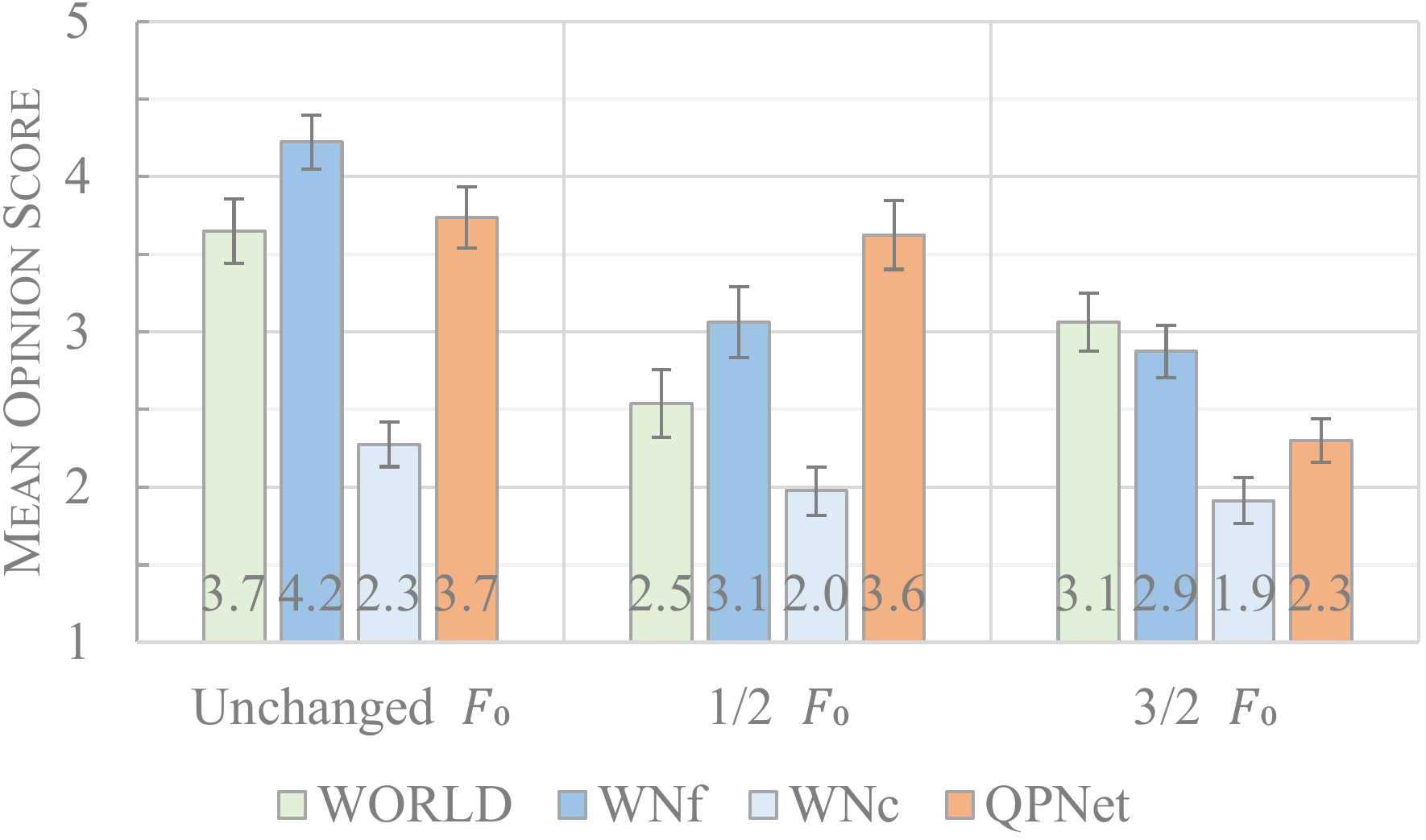}}
\caption{Sound quality MOS evaluation of female speakers with 95~\% CI.}
\label{fig:mos_female}
\end{figure}

\begin{figure}[t]
\centering
\centerline{\includegraphics[width=0.92\columnwidth]{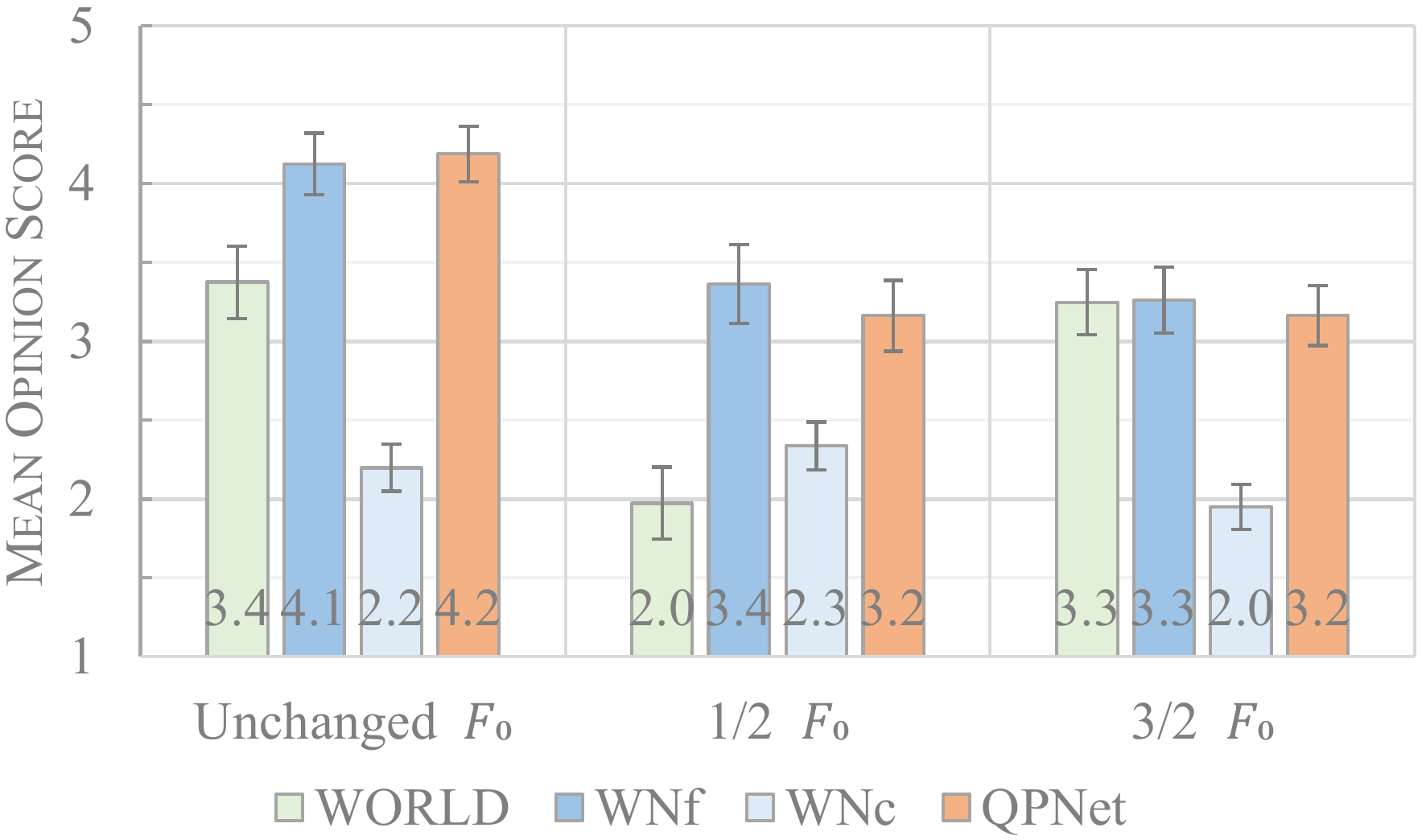}}
\caption{Sound quality MOS evaluation of male speakers with 95~\% CI.}
\label{fig:mos_male}
\end{figure}

\begin{figure}[t]
\centering
\centerline{\includegraphics[width=1.0\columnwidth]{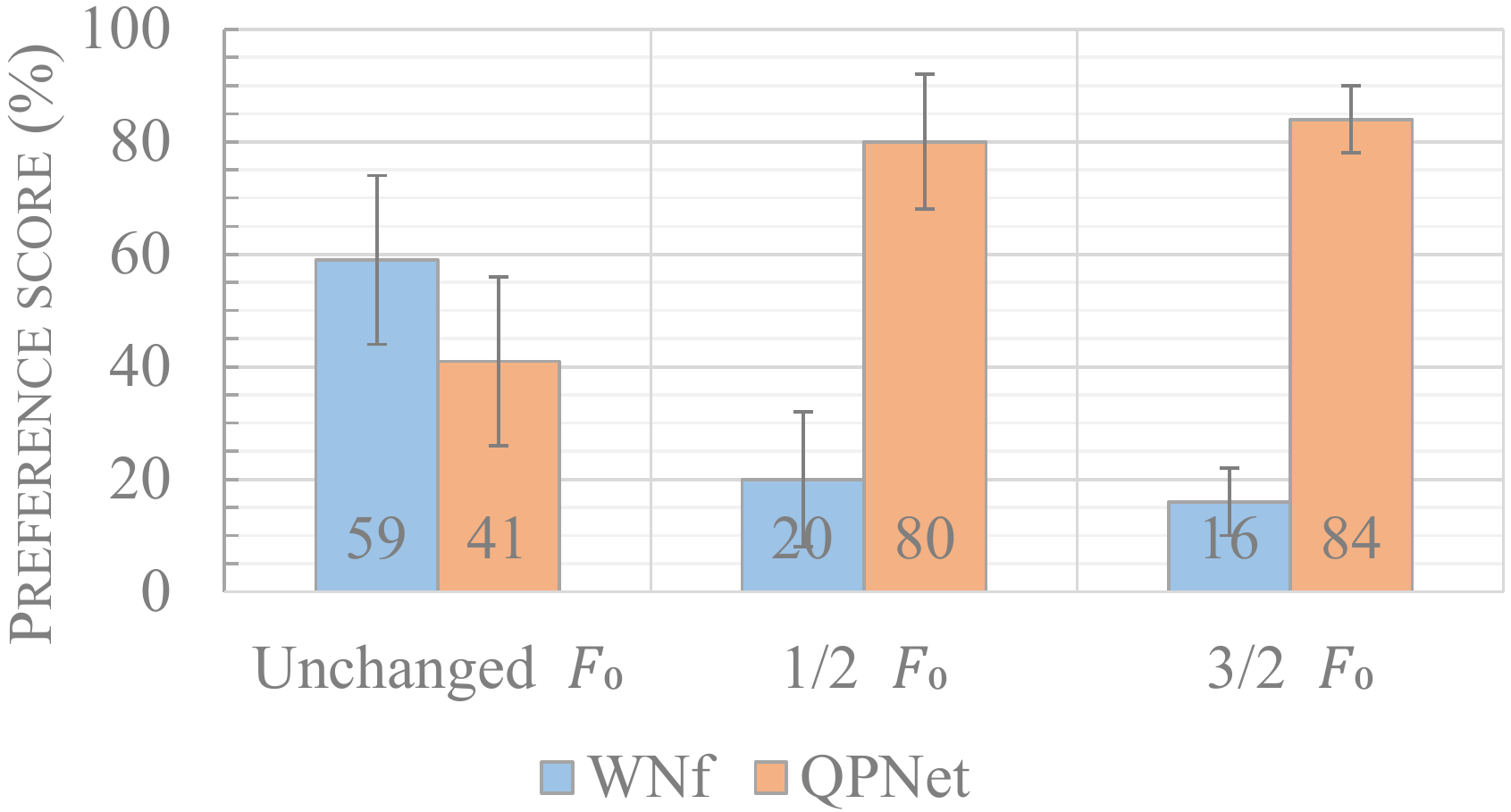}}
\caption{Pitch accuracy ABX evaluation of female speakers with 95~\% CI.}
\label{fig:abx_female}
\end{figure}

\begin{figure}[t]
\centering
\centerline{\includegraphics[width=1.0\columnwidth]{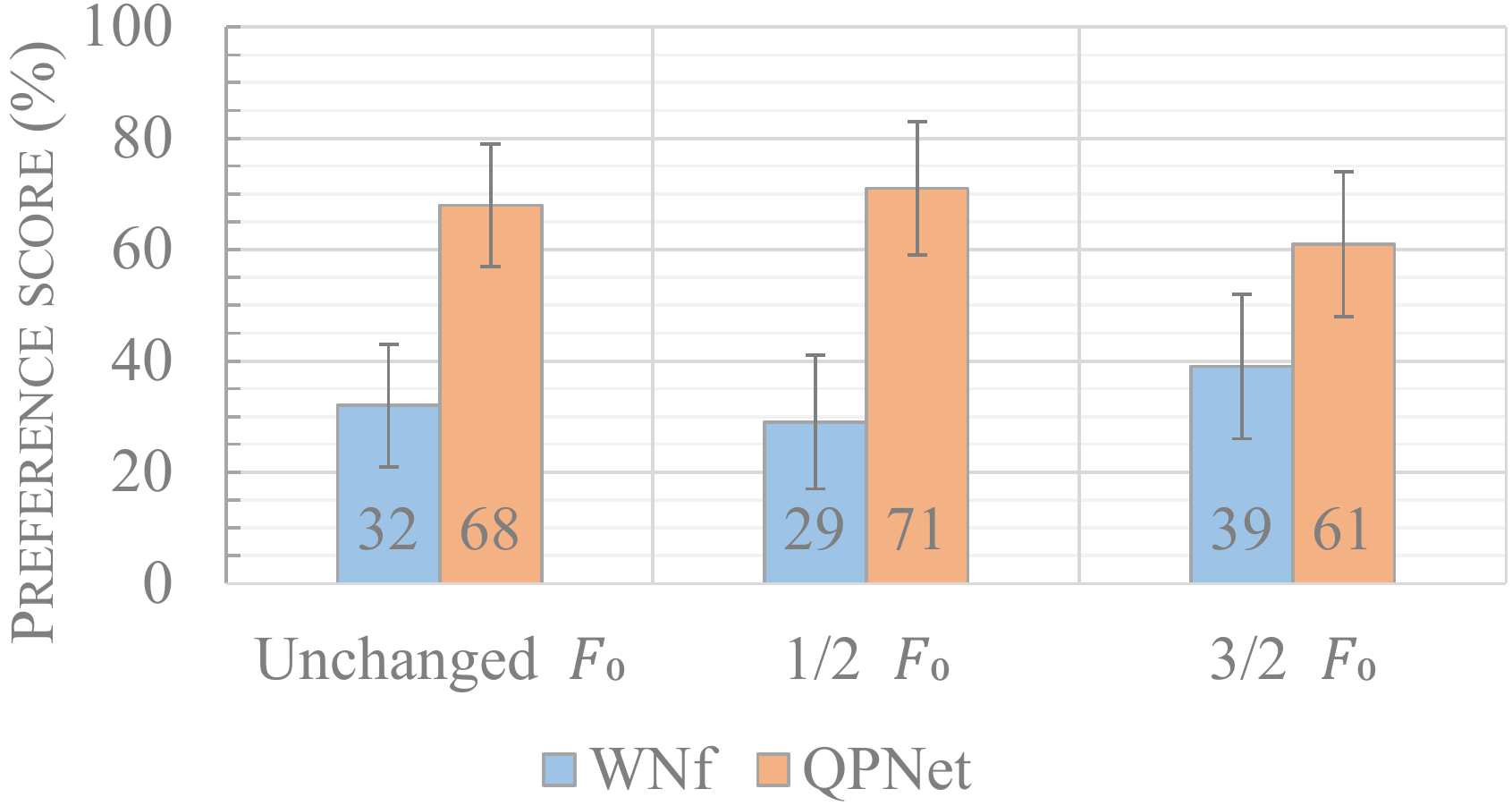}}
\caption{Pitch accuracy ABX evaluation of male speakers with 95~\% CI.}
\label{fig:abx_male}
\end{figure}

\subsubsection{Speech Quality}
As shown in Fig.~\ref{fig:mos_female}, for the female speaker set, the QPNet vocoder significantly outperforms the same-size WNc vocoder in all cases. Although the QPNet vocoder achieves slightly lower naturalness than the WNf vocoder in the unchanged $F_0$ (inside) case, the QPNet vocoder still attains markedly better naturalness than the WNf vocoder in the 1/2 $F_0$ (outside) case. The results indicate that halving the network size markedly degrades the speech modeling capability of the WN vocoder. However, the proposed PDCNNs significantly improves the modeling capacity with the halved network size, especially in the 1/2 $F_0$ case which makes QPNet obtain a long {\it effective receptive field} length. On the other hand, owing to the small dilated factors caused by the high $F_0$ values, many of the PDCNNs may degenerate to DCNNs in the 3/2 $F_0$ case. Specifically, when the dilated factors are less than or equal to one because of the high $F_0$ values, the dilation sizes of PDCNN are also less than or equal to DCNN. As a result, while these vocoders are conditioned on the auxiliary features with 3/2 $F_0$, although the QPNet vocoder still outperforms the WNc vocoder, the naturalness of the WNf- and WORLD-generated utterances is higher than that of the QPNet-generated utterances because of the much shorter {\it effective receptive field} length of the QPNet vocoder.

In addition, as the results of the male speaker set shown in Fig.~\ref{fig:mos_male}, the naturalness of the QPNet-generated utterances is comparable to that of the WNf-generated utterances and significantly better than that of the WNc-generated utterances in all $F_0$ cases. Specifically, even if the $F_0$ values are scaled, most of the 3/2 $F_0$ values of the male utterances are still within the range of the normal female $F_0$. Therefore, the {\it effective receptive field} lengths of the QPNet vocoder are still much longer than the {\it receptive field} lengths of the WNc vocoder for most male utterances with scaled $F_0$. On the other hand, the WORLD vocoder shows a similar tendency in the evaluations of both female and male speaker sets. In the unchanged $F_0$ case, the naturalness of the WORLD-generated utterances is slightly lower than the WNf- and QPNet-generated utterances. In the scaled $F_0$ cases, the WORLD vocoder achieves even much lower naturalness in the 1/2 $F_0$ case, but comparable naturalness in the 3/2 $F_0$ case.

\subsubsection{Pitch Accuracy and Summary}
As shown in Figs.~\ref{fig:abx_female} and~\ref{fig:abx_male}, the QPNet vocoder significantly outperforms the WNf vocoder in terms of pitch accuracy in most $F_0$ cases and both the female and male sets except in the unchanged $F_0$ cases of the female set, which may be caused by the naturalness degradation. The results confirm the pitch controllability improvement of the QPNet vocoder with the PDCNNs. In summary, the QPNet vocoder with the more compact network size achieves comparable speech quality to the WNf vocoder under most conditions except for the female set with 3/2 $F_0$ because the higher $F_0$ values may make the PDCNNs degenerate to the DCNNs. The QPNet vocoder conditioned on the unseen $F_0$ also gets the markedly higher pitch accuracy than the WNf vocoder. Moreover, the QPNet vocoder achieved higher or comparable speech quality than the WORLD vocoder under most conditions except conditioned on the acoustic features with the unseen 3/2 female $F_0$.

\subsection{Discussion}
\begin{figure}[t]
\centering
\centerline{\includegraphics[width=1.0\columnwidth]{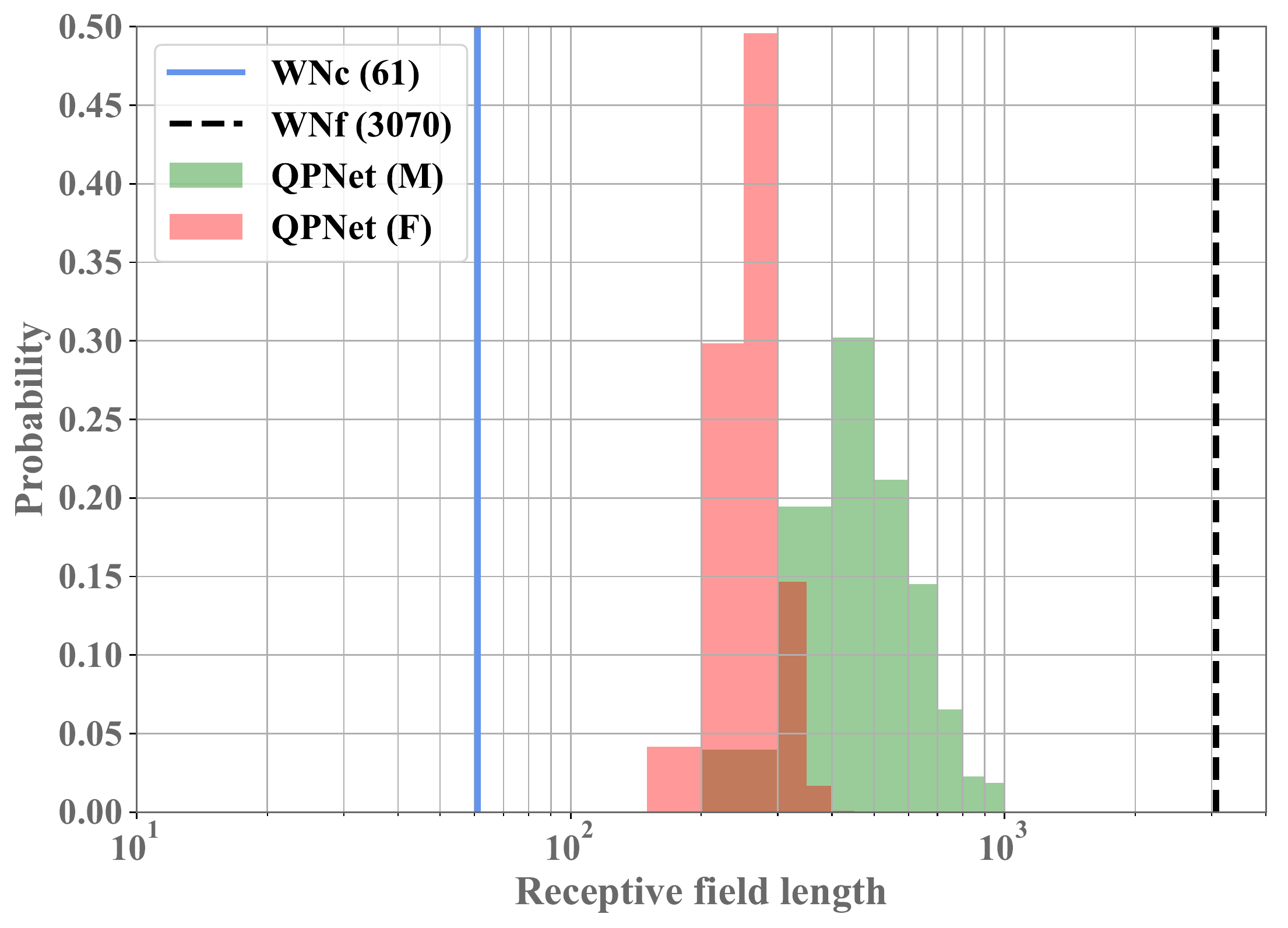}}
\caption{Distributions of {\it receptive field} lengths of different vocoders.}
\label{fig:rp_pmf}
\end{figure}

As shown in Fig.~\ref{fig:rp_pmf}, the length of the {\it receptive fields} of WNf is 3070 samples (The {\it receptive field} length of 10 blocks in each chunk is $2^0+2^1+\dots+2^9=1023,$ so the total length is $1023\times3$ with an extra one from the causal layer), that of WNc is 61 samples (Each chunk contains $2^0+2^1+2^2+2^3=15,$ so the total {\it receptive field} length is $15\times4+1=61$), and that of QPNet is 100--1000 samples (The {\it receptive field} length of the fixed blocks and the causal layer is $15\times3+1=46$, and that of the adaptive blocks is $15\times E_t$. The pitch-dependent dilated factor $E_t$ with a {\it dense factor} 8 was around 60 for 50~Hz and 6 for 500~Hz). Specifically, the {\it receptive field} lengths of WNf and WNc are constant because of the fixed network structure, and the {\it receptive field} length of QPNet is time-variant and pitch-dependent because of the QP structure. 

Furthermore, the results in Fig.~\ref{fig:rp_pmf} also show that the QPNet {\it effective receptive field} lengths of both SPOKE female and male speakers are longer than the {\it receptive field} length of WNc, which are consistent with the evaluation results showing that QPNet significantly outperforms WNc. Furthermore, most of the {\it effective receptive field} lengths of the female set are shorter than that of the male set, and it is caused by the higher $F_0$ values of the female speakers. The distribution results also imply that the {\it effective receptive field} lengths of QPNet are close to the {\it receptive field} length of WNc when conditioned on the female 3/2 $F_0$ because most PDCNNs degenerate to DCNNs. In conclusion, the performance of AR models is highly related to the length of the {\it receptive fields}.

However, the length of the {\it receptive fields} may be more strongly correlated to the quality of the generated speech, whereas a balanced proportion of the adaptive and fixed modules may be an essential factor for the pitch accuracy. Specifically, although the full-size QPNet has the longest {\it effective receptive field} lengths and achieves the lowest MCD, the pitch accuracy of full-size QPNet is still lower than that of compact-size QPNet. The possible reason is that the full-size QPNet is dominated by the fixed blocks because the number of the fixed blocks is much larger than the number of the adaptive blocks while the compact-size QPNet has more balanced numbers of the fixed and adaptive blocks.

In addition, as shown in Tables~\ref{tb:model1} and~\ref{tb:model2}, the number of the trainable parameters of the compact-size QPNet model is around half of that of the WNf model, so only about 75~\% of the training time and 40~\% of the generation time were required. However, because of the very long {\it effective receptive fields}, the memory usage of QPNet in the training stage was almost the same as that of WNf. The huge memory requirement in the training process limits the possible ratio of the fixed to adaptive modules, which leads to an unbalanced proportion problem. Therefore, improving the efficiency of memory usage will be one of the main tasks of future work.

\section{Conclusion} \label{conclusion}
In this paper, we propose a WaveNet-like audio waveform generation model named QPNet, which models quasi-periodic and high-temporal-resolution audio signals on the basis of an NN-based AR model with a novel PDCNN component and a cascaded AR structure. Specifically, the novel PDCNN component is a variant of a DCNN that dynamically changes the dilation size corresponding to the conditioned $F_0$ for modeling the long-term correlations of audio samples. On the basis of the sinusoid generation evaluation results, the PDCNNs improves the periodicity-modeling capability of the generation network using the introduced prior frequency information. Furthermore, the QPNet vocoder models the short- and long-term correlations of speech samples on the basis of the cascaded fixed and adaptive macroblocks, respectively. 

The speech generation evaluation results indicate that the proposed QPNet vocoder attains a higher pitch accuracy and comparable speech quality to the WN vocoder especially when conditioning on the unseen auxiliary $F_0$ values. Moreover, the network size and generation time requirements of the QPNet vocoder are only half of those of the WN vocoder. In conclusion, the proposed QPNet model with the novel PDCNN component and compact cascaded network architecture improves the pitch controllability of the vanilla WN model, and it makes the QPNet vocoder more in line with the definition of a vocoder. However, because the $F_0$-transformed ground-truth utterances are absent, the evaluation results might include some unknown biases. Therefore, in our future work, we plan to design a better evaluation scheme as well as to further improve the performance of our QPNet vocoder.


\bibliography{mybib}
\bibliographystyle{IEEEtran}
%

\begin{IEEEbiography}
[{\includegraphics[width=1in,height=1.25in,clip,keepaspectratio]{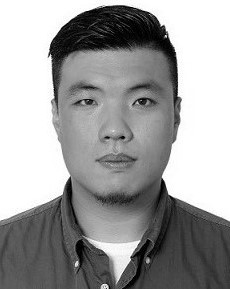}}]{Yi-Chiao Wu}
received his B.S and M.S degrees in engineering from the School of Communication Engineering of National Chiao Tung University in 2009 and 2011, respectively. He worked at Realtek, ASUS, and Academia Sinica for 5 years. Currently, he is pursuing his Ph.D. degree at the Graduate School of Informatics, Nagoya University. His research topics focus on speech generation applications based on machine learning methods, such as voice conversion and speech enhancement.
\end{IEEEbiography}

\begin{IEEEbiography}
[{\includegraphics[width=1in,height=1.25in,clip,keepaspectratio]{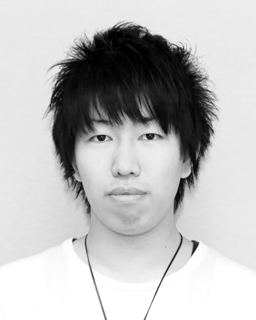}}]{Tomoki Hayashi}
received the B.E. degree in engineering and the M.E. and Ph.D. degrees in information science from Nagoya University, Japan, in 2014, 2016, and 2019, respectively. His research interests include statistical speech and audio signal processing. He is currently working as a postdoctoral researcher at Nagoya University and the chief operating officer of Human Dataware Lab. Co., Ltd. He received the IEEE SPS Japan 2020 Young Author Best Paper Award.
\end{IEEEbiography}


\begin{IEEEbiography}
[{\includegraphics[width=1in,height=1.25in,clip,keepaspectratio]{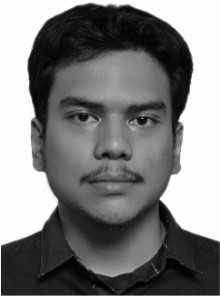}}]{Patrick Lumban Tobing}
received his B.E. degree from Bandung Institute of Technology (ITB), Indonesia, in 2014 and his M.E. degree from Nara Institute of Science and Technology (NAIST), Japan, in 2016. He completed his Ph.D. course at the Graduate School of Information Science, Nagoya University, Japan, in 2019, and is currently working as a Researcher. He received a Best Student Presentation Award from the Acoustical Society of Japan (ASJ). He is a member of IEEE and ISCA.
\end{IEEEbiography}

\begin{IEEEbiography}[{\includegraphics[width=1in,height=1.25in,clip,keepaspectratio]{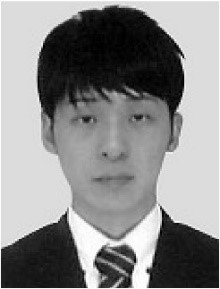}}]{Kazuhiro Kobayashi}
received his B.E. degree from the Department of Electrical and Electronic Engineering, Faculty of Engineering Science, Kansai University, Japan, in 2012, and his M.E. and Ph.D. degrees from Nara Institute of Science and Technology (NAIST), Japan, in 2014 and 2017, respectively. He is currently working as a Postdoctoral Researcher at the Graduate School of Information Science, Nagoya University, Japan. He has received a few awards including a Best Presentation Award from the Acoustical Society of Japan (ASJ). He is a member of IEEE, ISCA, and ASJ.
\end{IEEEbiography}

\begin{IEEEbiography}
[{\includegraphics[width=1in,height=1.25in,clip,keepaspectratio]{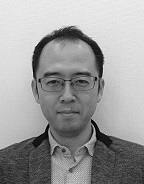}}]{Tomoki Toda}
is a Professor of the Information Technology Center at
Nagoya University, Japan. He received the B.E. degree from Nagoya
University in 1999, and the D.E. degree from the Nara Institute of
Science and Technology (NAIST), Japan, in 2003. He was a Research Fellow
of the Japan Society for the Promotion of Science from 2003 to 2005. He
was then an Assistant Professor (2005–2011) and an Associate Professor
(2011–2015) at NAIST. His research interests include statistical
approaches to speech, music, and environmental sound processing. He
received the IEEE SPS 2009 Young Author Best Paper Award and the 2013
EURASIP-ISCA Best Paper Award (Speech Communication Journal).
\end{IEEEbiography}




\end{document}